\documentclass[aps,prl,twocolumn,superscriptaddress]{revtex4-1}
\usepackage{amsmath}	
\usepackage{graphicx}	
\usepackage{color}
\usepackage{gensymb}
\usepackage{braket}
\usepackage{hyperref}
\usepackage{upgreek}
\usepackage[normalem]{ulem}

\begin{document}
	
\title{Interplay between spin proximity effect and charge-dependent exciton dynamics in MoSe$_2$ / CrBr$_3$ van der Waals heterostructures}
	
\author{T. P. Lyons}
\email{t.lyons@sheffield.ac.uk}
\author{D. Gillard}
\affiliation{Department of Physics and Astronomy, The University of Sheffield, Sheffield S3 7RH, UK}
\author{A. Molina-S\'anchez}
\affiliation{QuantaLab, International Iberian Nanotechnology Laboratory, 4715-330 Braga, Portugal}
\author{A. Misra}
\affiliation{School of Physics and Astronomy, The University of Manchester, Manchester M13 9PL, UK}
\affiliation{Department of Physics, Indian Institute of Technology Madras (IIT Madras), Chennai, India}
\author{F. Withers}
\affiliation{College of Engineering, Mathematics and Physical Sciences, University of Exeter, Exeter, EX4 4QF, UK}
\author{P. S. Keatley}
\affiliation{College of Engineering, Mathematics and Physical Sciences, University of Exeter, Exeter, EX4 4QF, UK}
\author{A. Kozikov}
\affiliation{School of Physics and Astronomy, The University of Manchester, Manchester M13 9PL, UK}
\author{T. Taniguchi}
\affiliation{National Institute for Materials Science, Tsukuba, Ibaraki 305-0044, Japan}
\author{K. Watanabe}
\affiliation{National Institute for Materials Science, Tsukuba, Ibaraki 305-0044, Japan}
\author{K. S. Novoselov}
\affiliation{School of Physics and Astronomy, The University of Manchester, Manchester M13 9PL, UK}
\affiliation{Centre for Advanced 2D Materials, National University of Singapore, 117546 Singapore}
\affiliation{Chongqing 2D Materials Institute, Liangjiang New Area, Chongqing, 400714, China}
\author{J. Fern\'andez-Rossier}
\affiliation{QuantaLab, International Iberian Nanotechnology Laboratory, 4715-330 Braga, Portugal}
\author{A. I. Tartakovskii}
\email{a.tartakovskii@sheffield.ac.uk}
\affiliation{Department of Physics and Astronomy, The University of Sheffield, Sheffield S3 7RH, UK}
	
\date{\today}

\begin{abstract}
	Semiconducting ferromagnet-nonmagnet interfaces in van der Waals heterostructures present a unique opportunity to investigate magnetic proximity interactions dependent upon a multitude of phenomena including valley and layer pseudospins, moir\'e periodicity, or exceptionally strong Coulomb binding. Here, we report a charge-state dependency of the magnetic proximity effects between MoSe$_2$ and CrBr$_3$ in photoluminescence, whereby the valley polarization of the MoSe$_2$ trion state conforms closely to the local CrBr$_3$ magnetization, while the neutral exciton state remains insensitive to the ferromagnet. We attribute this to spin-dependent interlayer charge transfer occurring on timescales between the exciton and trion radiative lifetimes. Going further, we uncover by both the magneto-optical Kerr effect and photoluminescence a domain-like spatial topography of contrasting valley polarization, which we infer to be labyrinthine or otherwise highly intricate, with features smaller than 400 nm corresponding to our optical resolution. Our findings offer a unique insight into the interplay between short-lived valley excitons and spin-dependent interlayer tunnelling, while also highlighting MoSe$_2$ as a promising candidate to optically interface with exotic spin textures in van der Waals structures.
\end{abstract}
	
	
\maketitle

\section{Introduction}

Few of the materials which have underpinned solid-state physics research over the past century now lack a 2-dimensional crystalline analogue \cite{zhou2018library,haastrup2018computational}. This is keenly exemplified by recent discoveries of long range magnetic order persisting into the monolayer limit in 2-dimensional van der Waals materials \cite{gong2017discovery,huang2017layer,fei2018two,lee2016ising,kim2019micromagnetometry}. However, the true potential of layered crystals lies not simply in having access to a selection of atomically thin proxies for conventional materials, but rather the ability to easily stack and combine these different materials into arbitrarily designed van der Waals heterostructures \cite{geim2013van}. Not only are such composites free of dangling bonds and growth-related lattice matching constraints, but also exhibit cumulative and hybridized properties superior to those of the constituent layers, and as such represent the building blocks for future generations of nanoscale information technologies \cite{alexeev2019resonantly,jin2018imaging,cao2018unconventional,yu2017moire,hunt2013massive,sharpe2019emergent}.

In this context, the discovery of 2-dimensional magnets in particular realizes a step-change for the nascent field of wholly van der Waals based spintronics \cite{fei2018two,huang2018electrical,gong2019two,burch2018magnetism,seyler2018ligand,gibertini2019magnetic,jiang2018controlling,mak2019probing,zhong2020layer}. One such material currently experiencing a resurgence of interest is chromium tribromide (CrBr$_3$) \cite{ghazaryan2018magnon,kim2019micromagnetometry,ciorciaro2020observation,zhang2019direct,chen2019direct}, a layered ferromagnetic semiconductor with Curie temperature 37 K and magnetization $3.85 \mu_B$ per Cr atom along the easy $c$-axis \cite{tsubokawa1960magnetic}. Recent work suggests that, in stark contrast to few-layer CrI$_3$ and CrCl$_3$ flakes which both exhibit antiferromagnetic interlayer ordering \cite{huang2017layer,zhong2020layer,klein2019enhancement}, the interlayer exchange in exfoliated CrBr$_3$ is ferromagnetic, although can depend on stacking order \cite{kim2019micromagnetometry,ghazaryan2018magnon,ciorciaro2020observation,chen2019direct}.

The incorporation of this emerging family of magnetic materials with optically active transition metal dichalcogenides (TMDs) is expected to combine the advantageous chiral optical selection rules and spin-valley locking of TMDs with the highly correlated and field-responsive long range ordering inherent to magnets \cite{xu2014spin,zhong2017van,tong2019magnetic,burch2018magnetism}. Proximity effects between monolayer TMDs and van der Waals ferromagnets have so far manifested as enhanced valley Zeeman splitting and/or modifications to the photoluminescence (PL) intensity \cite{zhong2017van,ciorciaro2020observation,zhong2020layer,seyler2018valley}. Where changes in the circular polarization degree of PL are reported, significantly broadened linewidths preclude the extraction of useful spectral information, and individual excitonic states cannot be resolved \cite{zhong2017van}.

\begin{figure*}
	\center
	\includegraphics{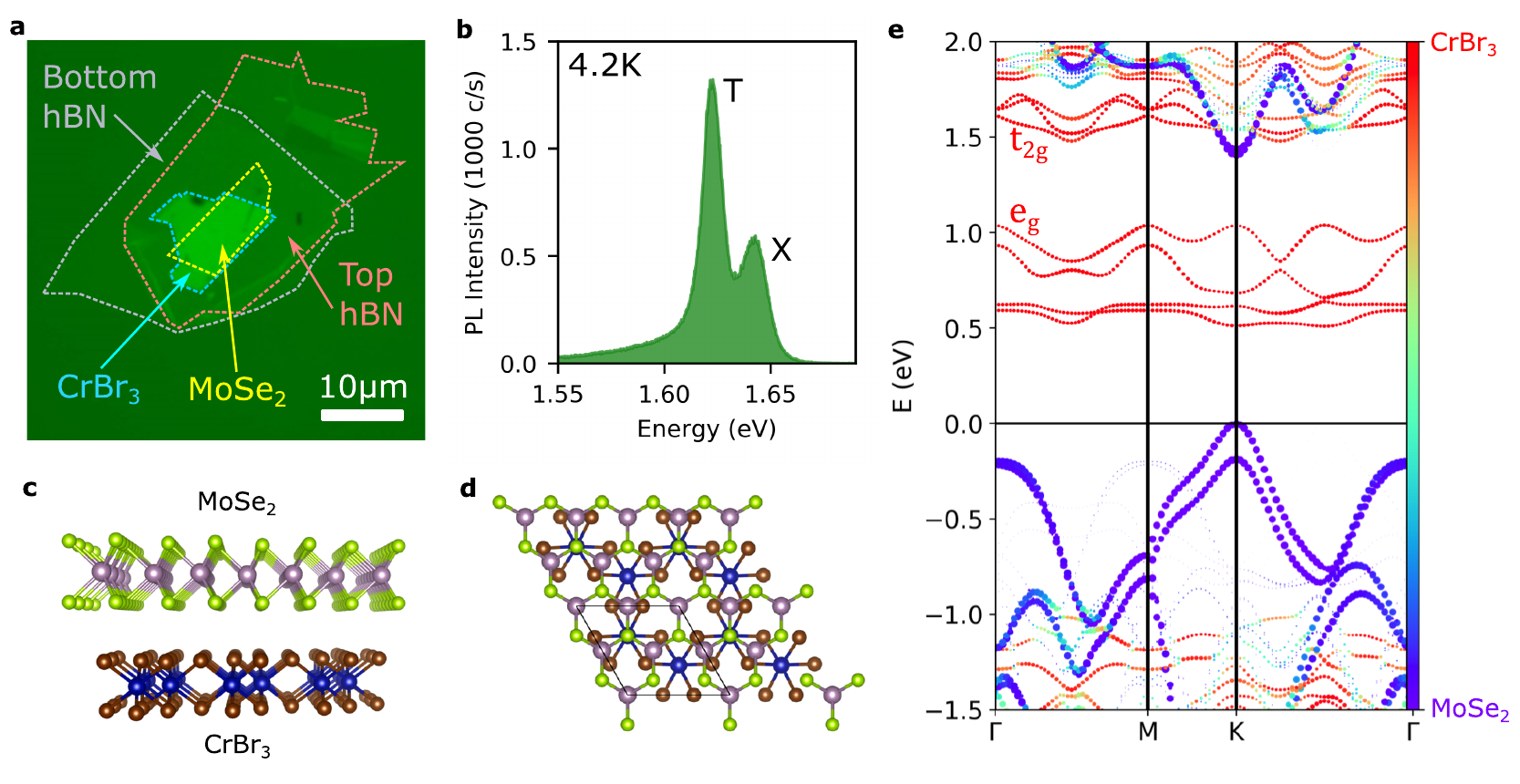}
	\caption{\label{fig:figure1} \textbf{MoSe$_2$ / CrBr$_3$ van der Waals heterostructures.} (a) Microscope image of the sample in this work showing multi-layered CrBr$_3$, monolayer MoSe$_2$, and hBN encapsulation layers. (b) Photoluminescence spectrum from the sample under continuous wave laser excitation at 1.946 eV and a sample temperature of 4.2 K. Neutral exciton (X) and trion (T) peaks are visible. (c,d) Schematics of a MoSe$_2$ / CrBr$_3$ heterobilayer structure used in the density functional theory calculations, viewed from the side (c) and top (d), where the supercell is highlighted. Here, the experimental lattice constant of monolayer MoSe$_2$ is applied to both layers, in order to simplify the calculations, along with setting the relative crystallographic twist angle to zero between the layers (see Supplementary Note 1). Crystal images produced using VESTA software \cite{momma2011vesta}. (e) DFT calculated electronic band structure of the MoSe$_2$ / CrBr$_3$ heterobilayer, projected on host material. The global valence band maximum is localized in MoSe$_2$ at $E=0$. Green band colour indicates hybridization between the two materials. The $e_g$-orbitals have spin polarization opposite to the $t_{2g}$-orbitals.}
\end{figure*} 

In this work, we study a van der Waals interface formed between the TMD molybdenum diselenide (MoSe$_2$) and ferromagnetic CrBr$_3$. While the interlayer band alignment is nominally type-II, with the global conduction band minimum found in CrBr$_3$, a strong spin polarization of the CrBr$_3$ conduction bands gives rise to spin-dependent interlayer charge transfer rates for electrons tunnelling from MoSe$_2$ to CrBr$_3$. This spin-dependent non-radiative decay channel causes a valley-dependent quenching of MoSe$_2$ photoluminescence intensity, leading to a strongly enhanced degree of circular polarization (DOCP). Crucially, the DOCP is observed only in the MoSe$_2$ trion state, rather than the neutral exciton, thereby providing insight into the interlayer tunnelling timescales, which we infer to lie between the exciton and trion radiative lifetimes. By performing wide-field polar Kerr microscopy, we both confirm that the MoSe$_2$ trion DOCP is an accurate indicator of local CrBr$_3$ magnetization, and gain information about the characteristic magnetic domain length-scales in our sample. Our findings shed new light on the mechanisms of interlayer charge transfer in TMD/magnet van der Waals interfaces. Moreover, our work highlights the importance of optical resolution when studying this new class of layered magnetic materials.
 
\section{Results}

\subsection{MoSe$_2$ / CrBr$_3$ van der Waals heterostructures}

\begin{figure*}
	\center
	\includegraphics{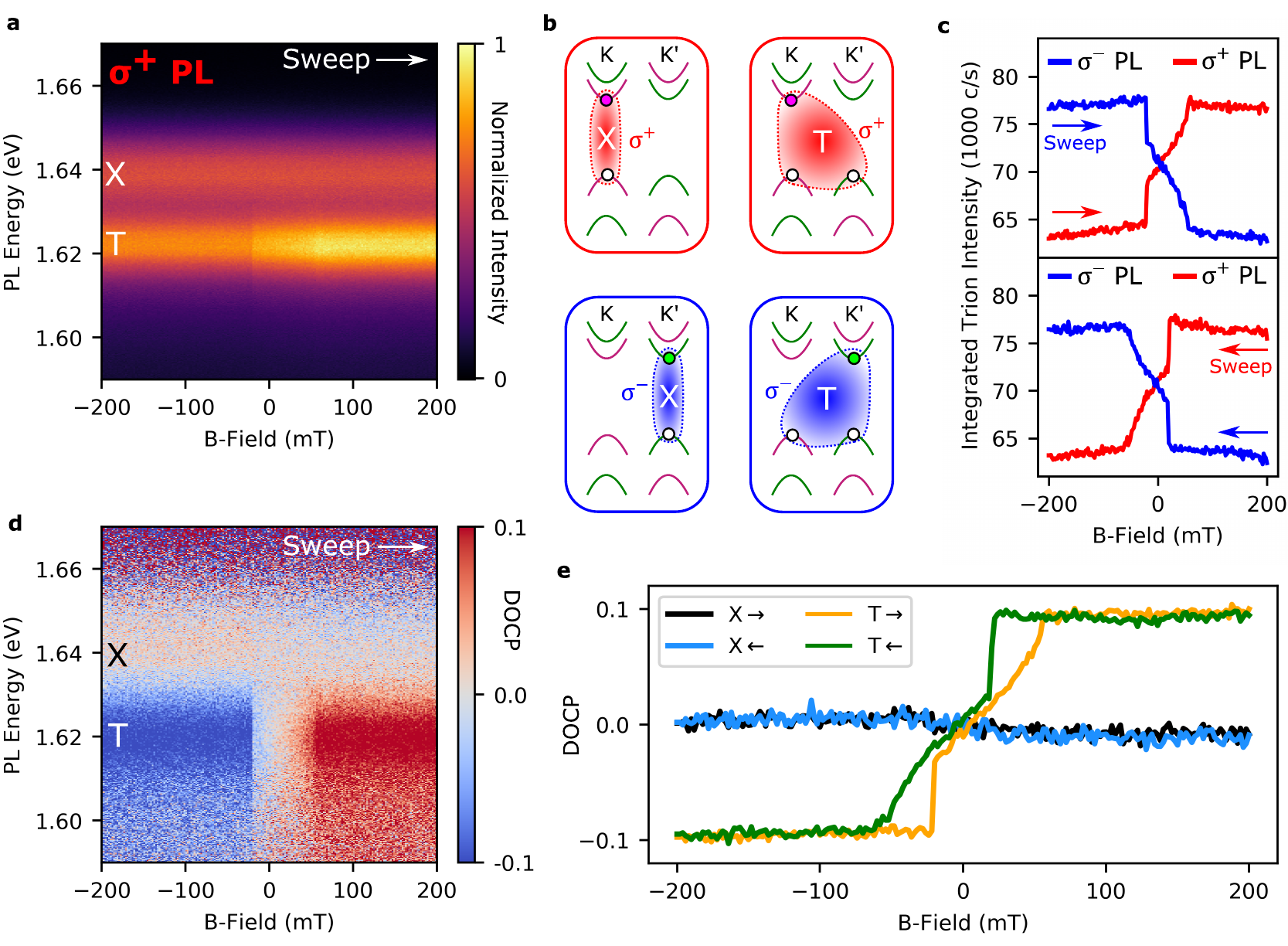}
	\caption{\label{fig:figure2} \textbf{Magneto-photoluminescence of MoSe$_2$ / CrBr$_3$ heterostructures.} (a) Photoluminescence (PL) intensity in $\sigma^+$ circular polarization as a function of photon energy and external B-field, in the forwards sweep direction. The neutral exciton (X) and trion (T) are labelled. (b) Schematics of the electron-hole valley configurations of the two ground state optically bright neutral excitons and positive trions in monolayer MoSe$_2$, which emit either a $\sigma^+$ or $\sigma^-$ polarized photon upon recombination. The pink (green) electron is spin $\uparrow$ ($\downarrow$), and the hole is white, regardless of spin. Spin-valley locking ties the electron spin to the emission polarization. (c) Integrated trion PL intensity in both circular polarizations, for the forward (upper panel) and backward (lower panel) B-field sweep directions. (d) Degree of circular polarization (DOCP) as a function of photon energy and B-field, in the forwards sweep direction. (e) DOCP of the exciton and trion states calculated from integrated intensities, shown for both sweep directions, revealing hysteresis type behaviour in the trion. The error in DOCP for this figure is $\sim \pm 0.005$.}
\end{figure*} 

The sample consists of a monolayer of MoSe$_2$ placed directly on top of $\sim 7 - 8$ nm thick multilayered CrBr$_3$, with few-layer hBN encapsulating the structure on both sides, as shown in Fig. 1a. The hBN encapsulation is necessary here owing to the extreme environmental sensitivity of exfoliated CrBr$_3$, a property shared also with CrI$_3$, in which degradation occurs rapidly under exposure to air and moisture, in a reaction catalysed by light \cite{shcherbakov2018raman}. The residue-free transfer techniques used during fabrication (see Methods) ensure full contact between the MoSe$_2$ and CrBr$_3$, such that trapped contaminants do not inhibit the interlayer charge transfer and proximity induced magnetic exchange effects. Photoluminescence (PL) from the sample at 4.2 K under non-resonant optical excitation at 1.946 eV displays a spectrum quite characteristic of MoSe$_2$, with a high energy peak ascribed to the neutral exciton (X) and a lower energy peak attributed to the charged exciton, or trion (T) (Fig. 1b) \cite{ross2013electrical,druppel2017diversity}. However, the emission intensity from the sample is far lower than typical for exfoliated MoSe$_2$, indicating a quenching of PL due to loss of photocarriers from MoSe$_2$ into CrBr$_3$, reducing the quantum yield.

Density functional theory (DFT) calculations of the electronic band structure of a MoSe$_2$ / CrBr$_3$ heterobilayer (Figs. 1c, d) indicate a type-II interlayer band alignment, with the global conduction band (CB) minimum located in the CrBr$_3$ layer, as shown in Fig. 1e. As the material band gaps depend sensitively on ligand electronegativity, the exact interlayer band alignment at the interface depends heavily on the elemental compositions of each layer \cite{molina2020magneto} (For details of the DFT calculations see Supplementary Note 1). Photogenerated electrons are predicted to scatter from the MoSe$_2$ CB into the CrBr$_3$ CB, introducing an additional non-radiative decay channel, corroborating with the observed suppressed luminescence intensity. Such interlayer charge transfer may be aided by electronic interlayer hybridization between the two materials, seen in Fig. 1e as green shaded regions. Any instances of hybridization would be expected to lead to efficient interlayer charge transfer, via elastic scattering events. Additionally, this heterostructure band alignment suggests that a static electron doping cannot exist in the MoSe$_2$ layer, while holes will accumulate in the MoSe$_2$ valence band (VB) as electrons tunnel out of the CB non-radiatively. Therefore, we attribute the observed trion luminescence to positively charged excitons, whereby neutral excitons bind to resident holes in the MoSe$_2$ VB \cite{ross2013electrical,zhong2017van}.

A crucial feature of our experiment is that our laser, at 1.946 eV, lies directly in an absorption minimum of CrBr$_3$ \cite{bermudez1979spectroscopic,dillon1966magneto,dillon1963magneto}. In this way, we ensure that photoexcitation occurs only in MoSe$_2$, and so no electron-hole pairs are created in CrBr$_3$. Therefore, with the exception of possible electrostatic discharge when the interface is first formed during sample fabrication, we can confidently neglect appreciable hole tunnelling from CrBr$_3$ to MoSe$_2$. As confirmation that our laser is not being absorbed, we search for direct CrBr$_3$ luminescence, reported at 1.35 eV under 1.77 eV illumination \cite{zhang2019direct}. Despite rigorous searching, we observe nothing but MoSe$_2$ PL between the laser energy and the limit of our detector sensitivity, $\sim 1.1$ eV, consistent with CrBr$_3$ being transparent to our laser.
	
The calculated electronic band structure in Fig. 1e also reveals a strong exchange splitting of the CrBr$_3$ CB, which is comprised of Cr$^{3+}$ $d$-orbitals. Bands arising from the majority-spin $e_{g}$ orbitals can be seen in red between 0.5 - 1 eV, while the minority-spin $t_{2g}$ orbitals form bands at 1.5 eV and above \cite{lado2017origin,gudelli2019magnetism}. Importantly, below the CrBr$_3$ Curie temperature $\sim 37$ K, these two sets of bands are oppositely spin polarized out of the sample plane. This is intrinsically tied to the local CrBr$_3$ magnetization, such that when the magnetization is flipped between pointing up or down out of plane, the spin polarization of these two sets of bands reverses. Since the $e_{g}$ orbitals constitute the majority-spin, they are spin-$\uparrow$ (spin-$\downarrow$) when the local CrBr$_3$ magnetization points upwards (downwards) out of the sample plane, while the $t_{2g}$ orbitals are spin-$\downarrow$ (spin-$\uparrow$).

Although our sample is not a heterobilayer, but rather multi-layered CrBr$_3$, we expect any magnetic proximity effects to arise primarily from the topmost CrBr$_3$ layer, owing to the very short range of exchange interactions and interlayer orbital wavefunction overlap \cite{zollnerexchange,zhong2020layer}. We confirm this by calculating and comparing the band structures of MoSe$_2$ / 1L CrBr$_3$ and MoSe$_2$ / 2L CrBr$_3$, and observe only negligible differences between the two (see Supplementary Note 2).

\subsection{Polarization resolved magneto-photoluminescence}

In order to gain insight into the valley pseudospin dynamics in our sample, we detect PL in $\sigma^+$ and $\sigma^-$ circular polarizations while sweeping an external B-field perpendicular to the sample plane, within the range B $=\pm 200$ mT, in both forward (negative to positive) and backward (positive to negative) sweep directions, at 4.2 K. A B-field orientation upwards (downwards) out of the plane is positive (negative). Fig. 2a shows the $\sigma^+$ polarized PL as a function of emitted photon energy and B-field, in the forward sweep direction. As can be seen, the trion intensity sharply increases at B $=-20$ mT, followed by a steady further increase, until the intensity reaches a plateau at B $\sim 50$ mT. This striking response must be attributed to the influence of CrBr$_3$, as the PL intensity of an isolated MoSe$_2$ monolayer would change only negligibly under the application of such weak external B-fields as used here \cite{li2014valley,wang2015magneto}. In stark contrast, the exciton emission, notably distinct from the trion, exhibits no observable change in intensity.

Bright neutral excitons are comprised of an electron-hole pair located in either the K or K' valley, corresponding to the $\sigma^+$ or $\sigma^-$ polarization state of the emitted photon (Fig. 2b) \cite{xu2014spin}. To form a positively charged trion, the exciton binds to an additional hole in the opposite valley (Fig. 2b) \cite{yu2015valley,ross2013electrical}. This intervalley ground state is highly favoured over the intravalley trion configuration, for which the large $\sim 200$ meV spin-orbit coupling strength in the VB is prohibitive \cite{yu2015valley,ross2013electrical,kormanyos2015k,Lyons2019}. The electron spin is tied to the emission helicity owing to optical selection rules and spin-valley locking \cite{xu2014spin,yu2015valley}.

\begin{figure*}
	\center
	\includegraphics{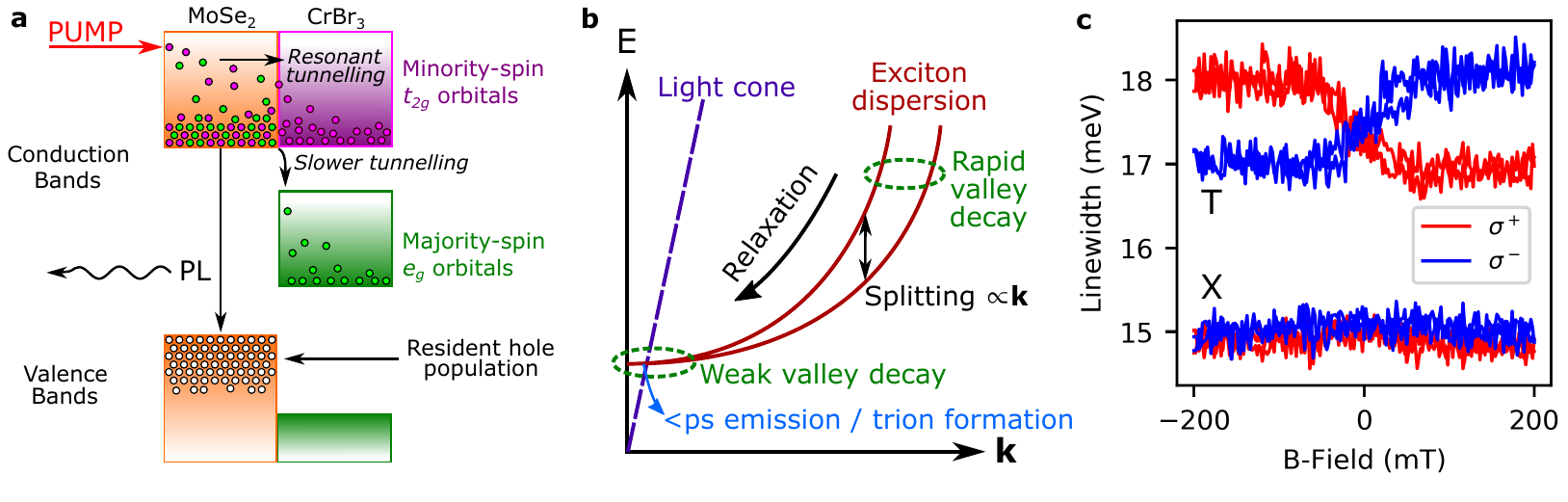}
	\caption{\label{fig:figure3} \textbf{Spin-dependent interlayer tunnelling.} (a) Schematic of spin-dependent interlayer charge transfer over the MoSe$_2$ / CrBr$_3$ interface. The type-II band alignment precludes a static electron doping in MoSe$_2$, and so we infer a resident hole population. In CrBr$_3$, the CB is strongly split into oppositely spin polarized bands. Only the minority-spin bands are resonant with the MoSe$_2$ CB. Upon optical injection of electron-hole pairs, the electrons may tunnel from MoSe$_2$ to CrBr$_3$, with an efficiency depending on their spin (green or purple indicate opposite spin). As CrBr$_3$ is transparent at our laser energy, absorption is negligible, and so we expect no appreciable dynamic hole transfer from CrBr$_3$ to MoSe$_2$. (b) Diagram of exciton dispersion in MoSe$_2$, with an energy splitting (the L-T splitting, see main text)  proportional to in-plane wavevector $\bf{k}$. The L-T splitting acts as an effective magnetic field inducing valley pseudospin relaxation in the presence of disorder scattering. Outside the light cone, efficient valley depolarization prevents spin-dependent tunnelling from influencing the eventual DOCP. At low $\bf{k}$, the spin-dependent tunnelling is able to influence the DOCP observed in PL. At vanishing centre of mass momentum, the exciton population either radiatively recombines (sub-ps) or else binds to a resident hole to form a long lived trion valley state, which will be susceptible to spin-dependent tunnelling. (c) Polarization resolved PL linewidths of the exciton (X) and trion (T). The trion linewidth shows opposing behaviour in different polarizations, indicating that the lifetime is influenced by spin-dependent interlayer tunnelling. The exciton has a constant linewidth, indicating insensitivity to CrBr$_3$ magnetization. }
\end{figure*} 

To elucidate the trion magneto-optical response, we plot the integrated trion intensity in both circular polarizations, in both forwards and backwards B-field sweep directions, as shown in Fig. 2c. As can be seen, the trion intensities in opposite polarizations behave exactly symmetrically, such that when the intensity of one polarization decreases the other increases by the same amount. In both sweep directions, the intensities invert as B passes through zero, indicating a clear reversal of trion valley populations when very small B-fields are applied. A specific response to the sweep direction which is symmetric about zero-field, as seen here, is a hallmark of ferromagnetism, confirming the influence of CrBr$_3$ on the MoSe$_2$ trion state valley polarization. 

A powerful tool to investigate the valley pseudospin dynamics in the sample is the DOCP, $(I_{\sigma^+} - I_{\sigma^-}) / (I_{\sigma^+} + I_{\sigma^-})$, of the emitted luminescence. Fig. 2d presents the DOCP as a function of photon energy and applied B-field, in the forward sweep direction. Here, a very prominent trion switching can be seen between robust positive and negative DOCP, while the exciton state remains apparently insensitive to any interactions with CrBr$_3$. Plotting the DOCP using integrated intensities, in both sweep directions, as shown in Fig. 2e, brings the sweep direction dependent response into sharp focus. While the exciton remains at zero, the trion state follows a sweep-dependent path when switching between the two valley polarized extremes, revealing the existence of two hysteretic lobes in the DOCP. We have fabricated and studied an additional 2 samples, both of which reproduce the trion DOCP sensitivity and exciton insensitivity observed in sample 1 (See Supplementary Note 3). Furthermore, by heating the sample under a fixed applied B-field, we observe the DOCP to tend to zero above the reported Curie temperature of CrBr$_3$, $\sim 37$ K (See Supplementary Note 4) \cite{tsubokawa1960magnetic}. 

\subsection{Spin-dependent interlayer charge transfer}

Ostensibly, an enhanced DOCP in MoSe$_2$ trion luminescence indicates a finite valley polarization in the resident carrier population. However, in this case, a hole valley polarization may be ruled out as the origin of the observed trion DOCP for three reasons. Firstly, any imbalance between the itinerant carrier populations of the two valleys leads not only to a change in trion intensity, but also to an opposite and proportional change in exciton intensity, owing to the intervalley nature of the trion ground state \cite{li2014valley}. We observe no corresponding change in exciton intensity in Fig. 2a, and no mirroring of the trion DOCP by the exciton DOCP in Figs 2d and 2e. Secondly, the observed trion DOCP cannot be due to a suppression of intervalley relaxation, as the sign of the DOCP is independent of the polarization state of the laser, and so the system retains no memory of optical valley initialization (see Supplementary Note 5). Thirdly, our DFT calculations predict a negligible proximity induced spin splitting in the MoSe$_2$ VB maxima, thereby precluding the generation of a static valley polarization in the itinerant hole population in thermal equilibrium. The band dependence of the proximity induced spin splitting is ultimately related to their different orbital content, with the VB orbitals being oriented preferably in the Molybdenum plane, and so experiencing a weak proximity effect. The orbitals constituting the MoSe$_2$ CB edges do protrude more out of the Molybdenum plane, and so do experience a modest spin splitting due to proximity, which we measure to be $< 0.5$ meV (Supplementary Notes 1 and 6). Regardless, this splitting is in the CB and so cannot generate a static hole valley polarization in the VB.

Therefore, rather than a static hole valley polarization, we attribute the observed DOCP to spin-dependent interlayer electron transfer, which translates to valley-dependent charge leakage from MoSe$_2$ owing to the spin-valley locking effect. Similar results have previously been observed in conventional ferromagnet-quantum well systems \cite{korenev2012dynamic}. The particular interlayer band alignment in this structure results in the MoSe$_2$ CB being resonant only with the minority-spin ($t_{2g}$) bands of CrBr$_3$, while the majority-spin ($e_g$) bands are non-resonant by several hundred meV (Fig. 1e). Therefore, we expect that photogenerated electrons in the MoSe$_2$ CB will be able to tunnel into the CrBr$_3$ CB more efficiently if they are spin aligned with the CrBr$_3$ minority-spin states. In order to tunnel into the majority-spin bands, multiple phonon scattering or high energy Auger-like interactions would be required, resulting in poor tunnelling efficiency (Fig. 3a). We also note the aforementioned lack of absorption of our laser by CrBr$_3$, which ensures that dynamic hole transfer (spin dependent or otherwise) from CrBr$_3$ into MoSe$_2$ can safely be neglected. Consequently, we can be confident that the observed trion DOCP arises purely from electron tunnelling from MoSe$_2$ into CrBr$_3$, which introduces an additional non-radiative decay for K or K' valley trion states, depending on the CrBr$_3$ magnetization.

We note that our PL signal is comparably quenched in both $\sigma^+$ and $\sigma^-$ polarizations, yet shows a 10 \% DOCP in the trion state. We attribute this to a complex interplay between exciton valley dynamics and spin-dependent interlayer charge transfer in our sample. In MoSe$_2$, the exciton dispersion experiences an energy splitting between excitons with dipole moment parallel (longitudinal, L) or perpendicular (transverse, T) to their in-plane wavevector, $\bf{k}$, arising from the long range electron-hole exchange interaction \cite{glazov2014exciton,yu2014dirac}. This L-T splitting is linearly proportional to $\bf{k}$ outside the light cone, and manifests as an effective magnetic field acting on the exciton valley pseudospin, causing precession and associated valley depolarization, aided by disorder scattering \cite{glazov2014exciton,yu2014dirac,maialle1993exciton}. Consequently, valley depolarization is extremely efficient ($<0.1$ ps \cite{dufferwiel2017valley}) for high-$\bf{k}$ excitons, and suppressed for low-$\bf{k}$ excitons (Fig. 3b). As such, any valley polarization from interlayer tunnelling at higher energies (closer to the laser at 1.946 eV) is quickly erased, and the tunnelling serves only to uniformly quench both the $\sigma^+$ and $\sigma^-$ PL spectra. We expect the tunnelling to be very efficient at these higher energies owing to resonant interlayer band alignment and regions of hybridization evident in Fig. 1e between the MoSe$_2$ CB and CrBr$_3$ $t_{2g}$ states. This efficient hot electron tunnelling will indeed be highly spin polarized, but the comparably fast valley depolarization in MoSe$_2$ ensures that no signatures persist in the observed DOCP. 
	
We therefore believe that the measured DOCP is sensitive only to spin/valley dynamics in the energy range corresponding to vanishing exciton centre-of-mass momentum, close to or inside the light cone. At these low wavevectors, the exchange induced valley depolarization is weak, allowing spin-dependent tunnelling to influence the DOCP in luminescence. The fact that DOCP is observed only in the trion state implies that the additional non-radiative decay arising from charge transfer in this low energy range must be slower than the exciton radiative lifetime (sub-ps \cite{jakubczyk2016radiatively}), but faster than the lifetime of the trion state, which is an order of magnitude longer ($> 10$ ps) \cite{Godde2016}. Our conclusion is supported by analysis of the PL linewidths, revealing a magnetization dependent change in the trion linewidth of $\sim 1$ meV, while the exciton linewidth remains constant. This allows us to estimate that the additional non-radiative decay experienced by the trion owing to spin-dependent charge transfer at these lower energies is $\sim 0.7$ ps. The tunnelling efficiency is generally influenced by various factors, including band hybridization, and energy or momentum offsets \cite{alexeev2019resonantly,ovesen2019interlayer,merkl2019ultrafast,froehlicher2018charge,selig2019theory}. Our calculations (Fig. 1e) predict that at the bottom of the MoSe$_2$ CB there is a low degree of hybridization with CrBr$_3$ CB states, and the energy and momentum offsets for possible electron scattering are large. These factors can explain the relatively long charge transfer times deduced from the experiment. We note, that the tunnelling rates may vary over a large range, as was revealed in recent pump-probe experiments on twisted TMD heterobilayers, where sub-ps to 5 ps tunnelling times were observed, drastically dependent on the relative crystallographic twist angle between the layers \cite{merkl2019ultrafast}.

In stark contrast to MoSe$_2$, the valley pseudospin lifetime in WSe$_2$ exceeds the PL lifetime \cite{zhu2014exciton}, and so the final PL DOCP will be influenced by the full extent of spin-dependent interlayer tunnelling occurring between the laser energy and the PL energy. This explains why the reported DOCP in WSe$_2$ / CrI$_3$ structures is larger than our MoSe$_2$ case, while also explaining why we observe no dependence on laser polarization in our sample (Supplementary Note 5), whereas results from WSe$_2$ / CrI$_3$ structures depend fundamentally on the laser polarization \cite{zhong2017van}.

\begin{figure*}
	\center
	\includegraphics{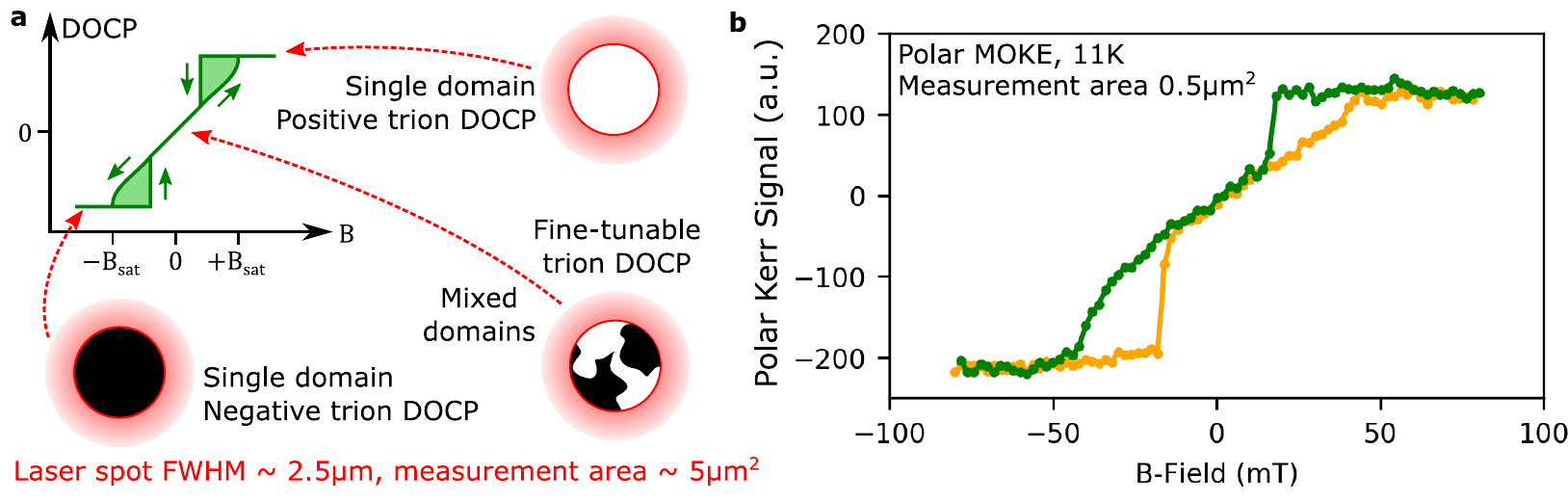}
	\caption{\label{fig:figure4} \textbf{Domain-like arrangement of trion valley polarization.} (a) Illustration of the trion degree of circular polarization (DOCP), with indications of the spatial arrangement of trion valley polarization at various points on the curve. Above $\lvert B_{sat} \rvert$, the CrBr$_3$ is in a single domain state and there is a uniform spin preference for electron tunnelling from MoSe$_2$ to CrBr$_3$. At lower B-fields, the relative strengths of spin $\uparrow$ or $\downarrow$ tunnelling may be fine tuned by control of the external field, which determines the relative spatial area occupied by adjacent domains. The red circles indicate an approximation of the laser spot area in the PL study, about 5 $\upmu$m$^2$, over which the domain behaviour is averaged. (b) Polar magneto-optical Kerr effect (MOKE) signal from the same sample, with a much smaller measurement area of 0.5 $\upmu$m$^2$. The domain nucleation behaviour is retained, indicating a very small length scale of domains. }
\end{figure*} 

\subsection{Domain-like valley polarization topography}

The intertwined relationship between trion PL and the local magnetization of CrBr$_3$, as demonstrated in Fig. 2e, means that the DOCP serves as a proxy magnetization or M-H curve. Viewed in this way, the trion polarization degree bears all the hallmarks of the magnetization response of a thin ferromagnetic film with perpendicular magnetocrystalline anisotropy, in which the magnetization of each domain must be aligned upwards or downwards out of the plane \cite{jagla2005hysteresis,zhang2019direct,kim2019micromagnetometry}. When $\lvert B \rvert > B_{sat} = 50$ mT, plateaus are observed, corresponding to magnetic saturation parallel to the external field, whereby the CrBr$_3$ is essentially in a single magnetic domain state. At field strengths lower than the saturation field, a complex response is observed featuring two marked hysteretic lobes, with a smooth linear gradient around B $=0$. Such a dependence is purely the result of magnetic domain dynamics, whereby the two discontinuities at B $= \pm 20$ mT appear as a result of spontaneous magnetic nucleation \cite{jagla2005hysteresis,zhang2019direct}. This process occurs when a saturating B-field is gradually removed, until a balance is tipped in favour of formation of antiparallel adjacent domains, representing a lower energy configuration. For the metastable saturation state to break down, a potential barrier associated with domain wall formation must be overcome \cite{jagla2005hysteresis}.

At very low fields, $\lvert B \rvert < 20$ mT, the average magnetization responds linearly to B owing to unopposed lateral domain wall motion through the flake, allowing domains to grow and shrink as appropriate. As the CrBr$_3$ flake is a single crystal, domain wall motion ought to be generally unobstructed, in contrast to polycrystalline thin film ferromagnets, in which grain boundaries may impede domain growth  \cite{riedi1985difference}. Introducing certain defects and disorder to the flake may also lead to domain wall pinning which opens up a hysteresis loop at zero field, consistent with the more typical remanence signatures of conventional ferromagnets \cite{zhong2017van,huang2017layer}. As such, the absence of coercivity in this sample indicates both a weak anisotropy \cite{ciorciaro2020observation}, and a low density of defects of the type which may impede lateral domain wall motion \cite{jeudy2018pinning,jagla2005hysteresis}. However, we note that there may be a high density of other defects which have a weaker influence on domain dynamics, consistent with the unstable nature of CrBr$_3$ (see Supplementary Note 8).

The crucial consequence of these CrBr$_3$ domain dynamics is that the spin-dependent interlayer charge transfer in the heterostructure adopts a spatial topography and magnetic response describable by the physics of ferromagnetic domains. As shown in Fig. 4a, when $ B  > B_{sat}$ ($ B  < -B_{sat}$), the charge transfer is preferentially of spin-$\downarrow$ (spin-$\uparrow$) over the entire interface. However, when $\lvert B \rvert < B_{sat}$, the interlayer charge transfer rates adopt a highly contrasting spatial arrangement over the sample, which in turn creates domain-like regions of opposite valley polarization co-existing in the MoSe$_2$ trion population. Cumulatively, reversal of the external B-field gives efficient selectivity between opposite preferential tunnelling rates for electrons of spin-$\uparrow$ or spin-$\downarrow$, while control of the field strength close to zero allows fine tuning of their relative prevalence.

We note that in our PL experiments, we measure an average of domain activity within the area of our laser spot, $\sim 5$ $\upmu$m$^2$. In order to gain insight into the characteristic length scales of the trion valley polarization domains in our sample, we measure the local magnetization via the polar magneto-optic Kerr effect (MOKE) in an optical microscope utilizing broadband illumination and wide-field high spatial resolution detection (Fig. 4b). We expect that the polar Kerr signal arises mainly from the CrBr$_3$ layers of the sample, given the maximum of the illumination energy at 2.7 eV closely matches the CrBr$_3$ main absorption resonance at 2.5 eV \cite{molina2020magneto}. Therefore, the close agreement between the PL from MoSe$_2$ and Kerr signal indicates a robust proximity effect and strong layer interaction \cite{catarina2019magneto}. Despite our MOKE measurement area being $\sim 0.5$ $\upmu$m$^2$, an order of magnitude smaller than in the PL experiment, there is close agreement between the two magneto-optical techniques, indicating that the domains are significantly smaller than the length scales of our optical probes. Indeed, no domain activity could be resolved in the Kerr microscope at all, confirming a domain size smaller than the optical resolution of the system, $\sim 400$ nm (See Supplementary Note 8). High spatial resolution techniques such as scanning probe magnetometry would be required to resolve the domain structure in our sample \cite{fei2018two,thiel2019probing}.

Our findings are in good agreement with previous work on CrBr$_3$ flakes in which electron microscopy reveals stripe domains with widths between $20-200$ nm \cite{matricardi1967electron}.
In thin film ferromagnets such as our CrBr$_3$ flake, domain size is governed by a complex interplay between magnetocrystalline and shape anisotropies, domain wall width, in-plane remanence, exchange energy, and magnetostatic interaction energy between top and bottom film surfaces \cite{virot2012theory}. The domain dynamics will therefore depend sensitively on the specific balance of these parameters, leading to significant variation. Indeed, complex and diverse domain patterns have been observed in CrBr$_3$ including stripe, honeycomb, bubble and cone domain structures \cite{kuhlow1975magnetic}. Weak in-plane external B-field orientation, disorder, temperature, or unpredictable local fluctuations, can all influence the domain patterns formed during each B-field cycle, with no correlation expected between patterns present before and after each instance of saturation, unless external input is used to provide control \cite{kuhlow1975magnetic}. We note that the trion Bohr radius in MoSe$_2$ ($\sim 1$ nm), is $1-2$ orders of magnitude smaller than the domain sizes inferred by our observations. This suggests that control of domain formation dynamics in CrBr$_3$, by the techniques mentioned above, would generate desired spatial patterns of contrasting electronic and trionic spin-valley polarization in MoSe$_2$, be they stripe, honeycomb, bubble, and so on, with important consequences for opto-spintronic device design at the nanoscale.

\subsection{Discussion}

To conclude, we observe spectrally dependent magnetic proximity effects in photoluminescence from monolayer MoSe$_2$ in contact with few-layer CrBr$_3$. The local out of plane magnetization in CrBr$_3$ is imprinted in the MoSe$_2$ trion state valley polarization, while the neutral exciton state bears no hallmarks of proximity interactions. We attribute the observed magneto-optical response to a spin-dependent interlayer charge transfer process, whereby electrons tunnel from the CB of MoSe$_2$ into the CB of CrBr$_3$ with differing rates depending on their real spin. This leaves the optically pumped exciton population in MoSe$_2$ subject to competing mechanisms of spin-dependent tunnelling and valley pseudospin depolarization, ultimately leading to the observed charge-state dependency of the proximity effects in photoluminescence. Furthermore, we infer that in the absence of a saturating applied B-field, the sample displays simultaneous manifestation of positive and negative valley polarization in the trion population, with an intricate spatial topography conforming to the magnetic domains of CrBr$_3$.

Our results demonstrate a highly localized ($< 400$ nm) technique to engineer the interlayer tunnelling rates of electrons in van der Waals heterostructures. The wide ranging implications of this extend to, for example, incorporation of a layered ferromagnet beneath a TMD heterobilayer to enable control of the interlayer exciton lifetime, or local spin and charge densities, serving as a powerful complimentary technique to design of the crystallographic twist angle. Such enticing avenues of band structure engineering are yet to be explored. A further tantalizing possibility is the mimicry of unconventional spin textures by 2-dimensional semiconductors in proximity to magnetic materials. The local sensitivity of excitonic valley polarization may be extended to employ monolayer MoSe$_2$ as an atomically thin optical probe for such exotic phenomena as skyrmions or domain wall racetrack devices, thus realizing a wealth of far reaching consequences along the boundary between exciton physics and van der Waals spintronics.

\section{Methods}

\subsection{Low temperature magneto-optical spectroscopy}
Low temperature magneto-photoluminescence spectroscopy was performed by mounting the sample in a liquid helium bath cryostat at 4.2 K with free space optical access and a superconducting magnet coil. Non-resonant continuous wave excitation at 1.946 eV in either $\sigma^{+}$ or $\sigma^{-}$ circular polarization was used, along with helicity selective circularly polarized PL collection, directed through a 0.75 m spectrometer and onto a nitrogen-cooled high sensitivity charge-coupled device. The helicity of the laser has no influence on the results measured, as discussed in Supplementary Note 5. To obtain linewidths and valley Zeeman splitting of exciton and trion peaks, Gaussian peak fitting was carried out on the PL spectra, as discussed in Supplementary Note 6.

\subsection{Density Functional Theory calculations}

See Supplementary Notes 1 and 2.

\subsection{Low temperature wide field Kerr microscopy}

Spatially resolved polar Kerr hysteresis loops were acquired at low temperature using a wide field Kerr microscope equipped with a small helium gas flow cryostat designed for microscopy. A long working distance, high numerical aperture (NA 0.7) objective lens was used to achieve a spatial resolution of ~400 nm. Spatial stability was achieved by mounting the cryostat on a piezoelectric stage for active drift correction. Hysteresis loops were acquired from user defined microscale regions of interest on the sample. The polar Kerr effect allowed sensitivity to the out-of-plane component of the magnetization in response to a magnetic field as it was swept perpendicular to the sample plane. The sample illumination was linearly polarized, while polarization changes of the reflected light due to the polar Kerr effect were detected as intensity changes using a high sensitivity CMOS camera after the light was passed through a nearly crossed analyzer. A background polarization rotation that was linear in applied field was measured from the sample substrate and subtracted from the acquired loops. The background was due to the Faraday rotation in the objective lens and cryostat window when subject to a field applied parallel to the polar axis. Differential imaging allowed non-magnetic contrast such as sample topography to be subtracted leaving only the contrast due to the out-of-plane component of the sample magnetization.

\subsection{Sample fabrication}

A heterostructure comprising of MoSe$_2$ / CrBr$_3$ encapsulated within thin hexagonal boron nitride (hBN) layers was assembled on a DBR substrate (top layer $\sim100$ nm SiO$_2$) following standard dry transfer procedures using a poly (methyl methacrylate) (PMMA) membrane. MoSe$_2$ crystals were exfoliated on 290 nm SiO$_2$ / Si substrate and monolayers of MoSe$_2$ were identified using optical contrast and room temperature PL measurements. To begin the heterostructure assembly, first, a monolayer of MoSe$_2$ was picked up by a thin hBN ($\sim 8 – 10$ nm) using the PMMA membrane. CrBr$_3$ crystals, purchased from commercial supplier HQ Graphene, were exfoliated on to a 290 nm SiO$_2$ / Si substrate in an argon-filled glove box maintaining the water and oxygen concentration less than 0.1 ppm. CrBr$_3$ flakes of different thicknesses (2-6 layers) could be easily identified using the contrast variation under different colour filters and dark-field imaging. However, for this work, a relatively thick $\sim$ 7-8 nm thick flake was used. A suitable (uniform and crack free) CrBr$_3$ layer was picked up by MoSe$_2$ / hBN / PMMA membrane in the glove box prepared in the first step. To complete the encapsulation of MoSe$_2$ / CrBr$_3$ between hBN layers, a thin hBN layer was further picked up by this membrane. Finally, the complete stack (hBN / CrBr$_3$ / MoSe$_2$ / hBN) on PMMA was dropped onto a DBR mirror by cutting the PMMA membrane. The PMMA was washed away in acetone and in IPA. 

\section{Acknowledgements}

T. P. L. acknowledges financial support from the EPSRC Doctoral Prize Fellowship scheme under Grant Reference EP/R513313/1. T. P. L. and A. I. T. acknowledge financial support from the European Graphene Flagship Project under grant agreement 785219, and EPSRC grants EP/P026850/1 and EP/S030751/1. The authors gratefully acknowledge the Exeter Time-Resolved Magnetism Facility (EXTREMAG - EPSRC Grant Reference EP/R008809/1) for the time allocated to this study for low temperature, wide field Kerr microscopy. The DFT calculations were performed on the Tirant III cluster of the Servei d`Inform\`atica of the University of Valencia (project vlc82) and on Mare Nostrum cluster of the Barcelona Supercomputing Center (project FI-2019-2-0034). A. M.-S. acknowledges the Marie-CurieCOFUND program Nano TRAIN For Growth II (Grant Agreement 713640). J. F.-R. acknowledges financial support from FCT for the grant UTAP-EXPL/NTec/0046/2017, as well as Generalitat Valenciana funding Prometeo 2017/139 and MINECO-Spain (Grant No. MAT2016-78625-C2). Growth of hexagonal boron nitride crystals was supported by the Elemental Strategy Initiative conducted by the MEXT, Japan and the CREST (JPMJCR15F3), JST.

\subsection{Correspondence}

Thomas Lyons (t.lyons@sheffield.ac.uk) or Alexander Tartakovskii (a.tartakovskii@sheffield.ac.uk).

\subsection{Data Availability}

Experimental data from this work is available by contacting the corresponding authors.

\color{black}

\bibliographystyle{ieeetr}
\bibliography{CrBr3Bib}

\begin{thebibliography}{10}

\bibitem{zhou2018library}
J.~Zhou, J.~Lin, X.~Huang, Y.~Zhou, Y.~Chen, J.~Xia, H.~Wang, Y.~Xie, H.~Yu,
  J.~Lei, {\em et~al.}, ``A library of atomically thin metal chalcogenides,''
  {\em Nature}, vol.~556, no.~7701, pp.~355--359, 2018.

\bibitem{haastrup2018computational}
S.~Haastrup, M.~Strange, M.~Pandey, T.~Deilmann, P.~S. Schmidt, N.~F. Hinsche,
  M.~N. Gjerding, D.~Torelli, P.~M. Larsen, A.~C. Riis-Jensen, {\em et~al.},
  ``The computational 2{D} materials database: high-throughput modeling and
  discovery of atomically thin crystals,'' {\em 2D Materials}, vol.~5, no.~4,
  p.~042002, 2018.

\bibitem{gong2017discovery}
C.~Gong, L.~Li, Z.~Li, H.~Ji, A.~Stern, Y.~Xia, T.~Cao, W.~Bao, C.~Wang,
  Y.~Wang, {\em et~al.}, ``Discovery of intrinsic ferromagnetism in
  two-dimensional van der {W}aals crystals,'' {\em Nature}, vol.~546, no.~7657,
  p.~265, 2017.

\bibitem{huang2017layer}
B.~Huang, G.~Clark, E.~Navarro-Moratalla, D.~R. Klein, R.~Cheng, K.~L. Seyler,
  D.~Zhong, E.~Schmidgall, M.~A. McGuire, D.~H. Cobden, {\em et~al.},
  ``Layer-dependent ferromagnetism in a van der {W}aals crystal down to the
  monolayer limit,'' {\em Nature}, vol.~546, no.~7657, p.~270, 2017.

\bibitem{fei2018two}
Z.~Fei, B.~Huang, P.~Malinowski, W.~Wang, T.~Song, J.~Sanchez, W.~Yao, D.~Xiao,
  X.~Zhu, A.~F. May, {\em et~al.}, ``Two-dimensional itinerant ferromagnetism
  in atomically thin {Fe}$_3${GeTe}$_2$,'' {\em Nature Materials}, vol.~17,
  no.~9, p.~778, 2018.

\bibitem{lee2016ising}
J.-U. Lee, S.~Lee, J.~H. Ryoo, S.~Kang, T.~Y. Kim, P.~Kim, C.-H. Park, J.-G.
  Park, and H.~Cheong, ``Ising-type magnetic ordering in atomically thin
  {FePS}$_3$,'' {\em Nano Letters}, vol.~16, no.~12, pp.~7433--7438, 2016.

\bibitem{kim2019micromagnetometry}
M.~Kim, P.~Kumaravadivel, J.~Birkbeck, W.~Kuang, S.~G. Xu, D.~Hopkinson,
  J.~Knolle, P.~A. McClarty, A.~Berdyugin, M.~B. Shalom, {\em et~al.},
  ``Micromagnetometry of two-dimensional ferromagnets,'' {\em Nature
  Electronics}, vol.~2, no.~10, pp.~457--463, 2019.

\bibitem{geim2013van}
A.~K. Geim and I.~V. Grigorieva, ``Van der {W}aals heterostructures,'' {\em
  Nature}, vol.~499, no.~7459, pp.~419--425, 2013.

\bibitem{alexeev2019resonantly}
E.~M. Alexeev, D.~A. Ruiz-Tijerina, M.~Danovich, M.~J. Hamer, D.~J. Terry,
  P.~K. Nayak, S.~Ahn, S.~Pak, J.~Lee, J.~I. Sohn, {\em et~al.}, ``Resonantly
  hybridized excitons in moir{\'e} superlattices in van der waals
  heterostructures,'' {\em Nature}, vol.~567, no.~7746, p.~81, 2019.

\bibitem{jin2018imaging}
C.~Jin, J.~Kim, M.~I.~B. Utama, E.~C. Regan, H.~Kleemann, H.~Cai, Y.~Shen,
  M.~J. Shinner, A.~Sengupta, K.~Watanabe, {\em et~al.}, ``Imaging of pure
  spin-valley diffusion current in {WS}$_2$-{WS}e$_2$ heterostructures,'' {\em
  Science}, vol.~360, no.~6391, pp.~893--896, 2018.

\bibitem{cao2018unconventional}
Y.~Cao, V.~Fatemi, S.~Fang, K.~Watanabe, T.~Taniguchi, E.~Kaxiras, and
  P.~Jarillo-Herrero, ``Unconventional superconductivity in magic-angle
  graphene superlattices,'' {\em Nature}, vol.~556, no.~7699, p.~43, 2018.

\bibitem{yu2017moire}
H.~Yu, G.-B. Liu, J.~Tang, X.~Xu, and W.~Yao, ``Moir{\'e} excitons: From
  programmable quantum emitter arrays to spin-orbit--coupled artificial
  lattices,'' {\em Science Advances}, vol.~3, no.~11, p.~e1701696, 2017.

\bibitem{hunt2013massive}
B.~Hunt, J.~Sanchez-Yamagishi, A.~Young, M.~Yankowitz, B.~J. LeRoy,
  K.~Watanabe, T.~Taniguchi, P.~Moon, M.~Koshino, P.~Jarillo-Herrero, {\em
  et~al.}, ``Massive {D}irac fermions and {H}ofstadter butterfly in a van der
  {W}aals heterostructure,'' {\em Science}, vol.~340, no.~6139, pp.~1427--1430,
  2013.

\bibitem{sharpe2019emergent}
A.~L. Sharpe, E.~J. Fox, A.~W. Barnard, J.~Finney, K.~Watanabe, T.~Taniguchi,
  M.~Kastner, and D.~Goldhaber-Gordon, ``Emergent ferromagnetism near
  three-quarters filling in twisted bilayer graphene,'' {\em Science},
  vol.~365, no.~6453, pp.~605--608, 2019.

\bibitem{huang2018electrical}
B.~Huang, G.~Clark, D.~R. Klein, D.~MacNeill, E.~Navarro-Moratalla, K.~L.
  Seyler, N.~Wilson, M.~A. McGuire, D.~H. Cobden, D.~Xiao, {\em et~al.},
  ``Electrical control of 2{D} magnetism in bilayer {CrI}$_3$,'' {\em Nature
  Nanotechnology}, vol.~13, no.~7, p.~544, 2018.

\bibitem{gong2019two}
C.~Gong and X.~Zhang, ``Two-dimensional magnetic crystals and emergent
  heterostructure devices,'' {\em Science}, vol.~363, no.~6428, p.~eaav4450,
  2019.

\bibitem{burch2018magnetism}
K.~S. Burch, D.~Mandrus, and J.-G. Park, ``Magnetism in two-dimensional van der
  {W}aals materials,'' {\em Nature}, vol.~563, no.~7729, pp.~47--52, 2018.

\bibitem{seyler2018ligand}
K.~L. Seyler, D.~Zhong, D.~R. Klein, S.~Gao, X.~Zhang, B.~Huang,
  E.~Navarro-Moratalla, L.~Yang, D.~H. Cobden, M.~A. McGuire, {\em et~al.},
  ``Ligand-field helical luminescence in a 2{D} ferromagnetic insulator,'' {\em
  Nature Physics}, vol.~14, no.~3, p.~277, 2018.

\bibitem{gibertini2019magnetic}
M.~Gibertini, M.~Koperski, A.~Morpurgo, and K.~Novoselov, ``Magnetic 2{D}
  materials and heterostructures,'' {\em Nature Nanotechnology}, vol.~14,
  no.~5, pp.~408--419, 2019.

\bibitem{jiang2018controlling}
S.~Jiang, L.~Li, Z.~Wang, K.~F. Mak, and J.~Shan, ``Controlling magnetism in
  2{D} {CrI}$_3$ by electrostatic doping,'' {\em Nature Nanotechnology},
  vol.~13, no.~7, p.~549, 2018.

\bibitem{mak2019probing}
K.~F. Mak, J.~Shan, and D.~C. Ralph, ``Probing and controlling magnetic states
  in 2{D} layered magnetic materials,'' {\em Nature Reviews Physics}, vol.~1,
  no.~11, pp.~646--661, 2019.

\bibitem{zhong2020layer}
D.~Zhong, K.~L. Seyler, X.~Linpeng, N.~P. Wilson, T.~Taniguchi, K.~Watanabe,
  M.~A. McGuire, K.-M.~C. Fu, D.~Xiao, W.~Yao, {\em et~al.}, ``Layer-resolved
  magnetic proximity effect in van der {W}aals heterostructures,'' {\em Nature
  Nanotechnology}, vol.~15, no.~3, pp.~187--191, 2020.

\bibitem{ghazaryan2018magnon}
D.~Ghazaryan, M.~T. Greenaway, Z.~Wang, V.~H. Guarochico-Moreira, I.~J.
  Vera-Marun, J.~Yin, Y.~Liao, S.~V. Morozov, O.~Kristanovski, A.~I.
  Lichtenstein, {\em et~al.}, ``Magnon-assisted tunnelling in van der {W}aals
  heterostructures based on {CrBr}$_3$,'' {\em Nature Electronics}, vol.~1,
  no.~6, p.~344, 2018.

\bibitem{ciorciaro2020observation}
L.~Ciorciaro, M.~Kroner, K.~Watanabe, T.~Taniguchi, and A.~Imamoglu,
  ``Observation of magnetic proximity effect using resonant optical
  spectroscopy of an electrically tunable {MoSe}$_2$ / {CrBr}$_3$
  heterostructure,'' {\em Physical Review Letters}, vol.~124, no.~19,
  p.~197401, 2020.

\bibitem{zhang2019direct}
Z.~Zhang, J.~Shang, C.~Jiang, A.~Rasmita, W.~Gao, and T.~Yu, ``Direct
  photoluminescence probing of ferromagnetism in monolayer two-dimensional
  {CrBr}$_3$,'' {\em Nano Letters}, vol.~19, no.~5, pp.~3138--3142, 2019.

\bibitem{chen2019direct}
W.~Chen, Z.~Sun, Z.~Wang, L.~Gu, X.~Xu, S.~Wu, and C.~Gao, ``Direct observation
  of van der waals stacking--dependent interlayer magnetism,'' {\em Science},
  vol.~366, no.~6468, pp.~983--987, 2019.

\bibitem{tsubokawa1960magnetic}
I.~Tsubokawa, ``On the magnetic properties of a {CrBr}$_3$ single crystal,''
  {\em Journal of the Physical Society of Japan}, vol.~15, no.~9,
  pp.~1664--1668, 1960.

\bibitem{klein2019enhancement}
D.~R. Klein, D.~MacNeill, Q.~Song, D.~T. Larson, S.~Fang, M.~Xu, R.~A. Ribeiro,
  P.~C. Canfield, E.~Kaxiras, R.~Comin, {\em et~al.}, ``Enhancement of
  interlayer exchange in an ultrathin two-dimensional magnet,'' {\em Nature
  Physics}, vol.~15, no.~12, pp.~1255--1260, 2019.

\bibitem{xu2014spin}
X.~Xu, W.~Yao, D.~Xiao, and T.~F. Heinz, ``Spin and pseudospins in layered
  transition metal dichalcogenides,'' {\em Nature Physics}, vol.~10, no.~5,
  pp.~343--350, 2014.

\bibitem{zhong2017van}
D.~Zhong, K.~L. Seyler, X.~Linpeng, R.~Cheng, N.~Sivadas, B.~Huang,
  E.~Schmidgall, T.~Taniguchi, K.~Watanabe, M.~A. McGuire, {\em et~al.}, ``Van
  der {W}aals engineering of ferromagnetic semiconductor heterostructures for
  spin and valleytronics,'' {\em Science Advances}, vol.~3, no.~5, p.~e1603113,
  2017.

\bibitem{tong2019magnetic}
Q.~Tong, M.~Chen, and W.~Yao, ``Magnetic proximity effect in a van der {W}aals
  moir{\'e} superlattice,'' {\em Physical Review Applied}, vol.~12, no.~2,
  p.~024031, 2019.

\bibitem{seyler2018valley}
K.~L. Seyler, D.~Zhong, B.~Huang, X.~Linpeng, N.~P. Wilson, T.~Taniguchi,
  K.~Watanabe, W.~Yao, D.~Xiao, M.~A. McGuire, {\em et~al.}, ``Valley
  manipulation by optically tuning the magnetic proximity effect in {WSe}$_2$ /
  {CrI}$_3$ heterostructures,'' {\em Nano Letters}, vol.~18, no.~6,
  pp.~3823--3828, 2018.

\bibitem{momma2011vesta}
K.~Momma and F.~Izumi, ``{VESTA} 3 for three-dimensional visualization of
  crystal, volumetric and morphology data,'' {\em Journal of Applied
  Crystallography}, vol.~44, no.~6, pp.~1272--1276, 2011.

\bibitem{shcherbakov2018raman}
D.~Shcherbakov, P.~Stepanov, D.~Weber, Y.~Wang, J.~Hu, Y.~Zhu, K.~Watanabe,
  T.~Taniguchi, Z.~Mao, W.~Windl, {\em et~al.}, ``Raman spectroscopy,
  photocatalytic degradation, and stabilization of atomically thin chromium
  tri-iodide,'' {\em Nano Letters}, vol.~18, no.~7, pp.~4214--4219, 2018.

\bibitem{ross2013electrical}
J.~S. Ross, S.~Wu, H.~Yu, N.~J. Ghimire, A.~M. Jones, G.~Aivazian, J.~Yan,
  D.~G. Mandrus, D.~Xiao, W.~Yao, {\em et~al.}, ``Electrical control of neutral
  and charged excitons in a monolayer semiconductor,'' {\em Nature
  Communications}, vol.~4, p.~1474, 2013.

\bibitem{druppel2017diversity}
M.~Dr{\"u}ppel, T.~Deilmann, P.~Kr{\"u}ger, and M.~Rohlfing, ``Diversity of
  trion states and substrate effects in the optical properties of an {MoS}$_2$
  monolayer,'' {\em Nature Communications}, vol.~8, no.~1, p.~2117, 2017.

\bibitem{molina2020magneto}
A.~Molina-S{\'a}nchez, G.~Catarina, D.~Sangalli, and J.~Fernandez-Rossier,
  ``Magneto-optical response of chromium trihalide monolayers: chemical
  trends,'' {\em Journal of Materials Chemistry C}, vol.~8, pp.~8856--8863,
  2020.

\bibitem{bermudez1979spectroscopic}
V.~M. Bermudez and D.~S. McClure, ``Spectroscopic studies of the
  two-dimensional magnetic insulators chromium trichloride and chromium
  tribromide—{I},'' {\em Journal of Physics and Chemistry of Solids},
  vol.~40, no.~2, pp.~129--147, 1979.

\bibitem{dillon1966magneto}
J.~Dillon~Jr, H.~Kamimura, and J.~Remeika, ``Magneto-optical properties of
  ferromagnetic chromium trihalides,'' {\em Journal of Physics and Chemistry of
  Solids}, vol.~27, no.~9, pp.~1531--1549, 1966.

\bibitem{dillon1963magneto}
J.~Dillon~Jr, H.~Kamimura, and J.~Remeika, ``Magneto-optical studies of
  chromium tribromide,'' {\em Journal of Applied Physics}, vol.~34, no.~4,
  pp.~1240--1245, 1963.

\bibitem{lado2017origin}
J.~L. Lado and J.~Fern{\'a}ndez-Rossier, ``On the origin of magnetic anisotropy
  in two dimensional {CrI}$_3$,'' {\em 2D Materials}, vol.~4, no.~3, p.~035002,
  2017.

\bibitem{gudelli2019magnetism}
V.~K. Gudelli and G.-Y. Guo, ``Magnetism and magneto-optical effects in bulk
  and few-layer {CrI}$_3$: a theoretical {GGA+ U} study,'' {\em New Journal of
  Physics}, vol.~21, no.~5, p.~053012, 2019.

\bibitem{zollnerexchange}
K.~Zollner, P.~E. Faria~Junior, and J.~Fabian, ``Proximity exchange effects in
  {MoSe}$_2$ and {WSe}$_2$ heterostructures with {CrI}$_3$: {T}wist angle,
  layer, and gate dependence,'' {\em Physical Review B}, vol.~100, p.~085128,
  Aug 2019.

\bibitem{li2014valley}
Y.~Li, J.~Ludwig, T.~Low, A.~Chernikov, X.~Cui, G.~Arefe, Y.~D. Kim, A.~M. Van
  Der~Zande, A.~Rigosi, H.~M. Hill, {\em et~al.}, ``Valley splitting and
  polarization by the {Z}eeman effect in monolayer {MoSe}$_2$,'' {\em Physical
  Review Letters}, vol.~113, no.~26, p.~266804, 2014.

\bibitem{wang2015magneto}
G.~Wang, L.~Bouet, M.~M. Glazov, T.~Amand, E.~L. Ivchenko, E.~Palleau,
  X.~Marie, and B.~Urbaszek, ``Magneto-optics in transition metal diselenide
  monolayers,'' {\em 2D Materials}, vol.~2, no.~3, p.~034002, 2015.

\bibitem{yu2015valley}
H.~Yu, X.~Cui, X.~Xu, and W.~Yao, ``Valley excitons in two-dimensional
  semiconductors,'' {\em National Science Review}, vol.~2, no.~1, pp.~57--70,
  2015.

\bibitem{kormanyos2015k}
A.~Korm{\'a}nyos, G.~Burkard, M.~Gmitra, J.~Fabian, V.~Z{\'o}lyomi, N.~D.
  Drummond, and V.~Fal’ko, ``\textbf{k{\textperiodcentered}p} theory for
  two-dimensional transition metal dichalcogenide semiconductors,'' {\em 2D
  Materials}, vol.~2, no.~2, p.~022001, 2015.

\bibitem{Lyons2019}
T.~P. Lyons, S.~Dufferwiel, M.~Brooks, F.~Withers, T.~Taniguchi, K.~Watanabe,
  K.~S. Novoselov, G.~Burkard, and A.~I. Tartakovskii, ``{The valley {Z}eeman
  effect in inter- and intra-valley trions in monolayer WSe$_2$},'' {\em Nature
  Communications}, vol.~10, no.~2330, 2019.

\bibitem{korenev2012dynamic}
V.~Korenev, I.~Akimov, S.~Zaitsev, V.~Sapega, L.~Langer, D.~Yakovlev, Y.~A.
  Danilov, and M.~Bayer, ``Dynamic spin polarization by orientation-dependent
  separation in a ferromagnet--semiconductor hybrid,'' {\em Nature
  Communications}, vol.~3, no.~1, pp.~1--7, 2012.

\bibitem{glazov2014exciton}
M.~Glazov, T.~Amand, X.~Marie, D.~Lagarde, L.~Bouet, and B.~Urbaszek, ``Exciton
  fine structure and spin decoherence in monolayers of transition metal
  dichalcogenides,'' {\em Physical Review B}, vol.~89, no.~20, p.~201302, 2014.

\bibitem{yu2014dirac}
H.~Yu, G.-B. Liu, P.~Gong, X.~Xu, and W.~Yao, ``Dirac cones and {D}irac saddle
  points of bright excitons in monolayer transition metal dichalcogenides,''
  {\em Nature Communications}, vol.~5, no.~1, pp.~1--7, 2014.

\bibitem{maialle1993exciton}
M.~Maialle, E.~d.~A. e~Silva, and L.~Sham, ``Exciton spin dynamics in quantum
  wells,'' {\em Physical Review B}, vol.~47, no.~23, p.~15776, 1993.

\bibitem{dufferwiel2017valley}
S.~Dufferwiel, T.~P. Lyons, D.~D. Solnyshkov, A.~A.~P. Trichet, F.~Withers,
  S.~Schwarz, G.~Malpuech, J.~M. Smith, K.~S. Novoselov, M.~S. Skolnick, {\em
  et~al.}, ``Valley-addressable polaritons in atomically thin semiconductors,''
  {\em Nature Photonics}, vol.~11, no.~8, p.~497, 2017.

\bibitem{jakubczyk2016radiatively}
T.~Jakubczyk, V.~Delmonte, M.~Koperski, K.~Nogajewski, C.~Faugeras,
  W.~Langbein, M.~Potemski, and J.~Kasprzak, ``Radiatively limited dephasing
  and exciton dynamics in {MoSe}$_2$ monolayers revealed with four-wave mixing
  microscopy,'' {\em Nano Letters}, vol.~16, no.~9, pp.~5333--5339, 2016.

\bibitem{Godde2016}
T.~Godde, D.~Schmidt, J.~Schmutzler, M.~A{\ss}mann, J.~Debus, F.~Withers, E.~M.
  Alexeev, O.~{Del Pozo-Zamudio}, O.~V. Skrypka, K.~S. Novoselov, M.~Bayer, and
  A.~I. Tartakovskii, ``{Exciton and trion dynamics in atomically thin
  {MoSe}$_2$ and {WSe}$_2$ : {E}ffect of localization},'' {\em Physical Review
  B}, vol.~94, p.~165301, oct 2016.

\bibitem{ovesen2019interlayer}
S.~Ovesen, S.~Brem, C.~Linder{\"a}lv, M.~Kuisma, T.~Korn, P.~Erhart, M.~Selig,
  and E.~Malic, ``Interlayer exciton dynamics in van der {W}aals
  heterostructures,'' {\em Communications Physics}, vol.~2, no.~1, pp.~1--8,
  2019.

\bibitem{merkl2019ultrafast}
P.~Merkl, F.~Mooshammer, P.~Steinleitner, A.~Girnghuber, K.-Q. Lin, P.~Nagler,
  J.~Holler, C.~Sch{\"u}ller, J.~M. Lupton, T.~Korn, {\em et~al.}, ``Ultrafast
  transition between exciton phases in van der {W}aals heterostructures,'' {\em
  Nature Materials}, vol.~18, no.~7, pp.~691--696, 2019.

\bibitem{froehlicher2018charge}
G.~Froehlicher, E.~Lorchat, and S.~Berciaud, ``Charge versus energy transfer in
  atomically thin graphene-transition metal dichalcogenide van der {W}aals
  heterostructures,'' {\em Physical Review X}, vol.~8, no.~1, p.~011007, 2018.

\bibitem{selig2019theory}
M.~Selig, E.~Malic, K.~J. Ahn, N.~Koch, and A.~Knorr, ``Theory of optically
  induced {F\"o}rster coupling in van der {W}aals coupled heterostructures,''
  {\em Physical Review B}, vol.~99, no.~3, p.~035420, 2019.

\bibitem{zhu2014exciton}
C.~Zhu, K.~Zhang, M.~Glazov, B.~Urbaszek, T.~Amand, Z.~Ji, B.~Liu, and
  X.~Marie, ``Exciton valley dynamics probed by {K}err rotation in {WSe}$_2$
  monolayers,'' {\em Physical Review B}, vol.~90, no.~16, p.~161302, 2014.

\bibitem{jagla2005hysteresis}
E.~A. Jagla, ``Hysteresis loops of magnetic thin films with perpendicular
  anisotropy,'' {\em Physical Review B}, vol.~72, no.~9, p.~094406, 2005.

\bibitem{riedi1985difference}
P.~Riedi and I.~Zammit-Mangion, ``Difference between domain wall motion in
  single crystal and polycrystalline {YIG} observed by pulsed {NMR},'' {\em
  Physica Status Solidi. A, Applied Research}, vol.~87, no.~2, pp.~K163--K166,
  1985.

\bibitem{jeudy2018pinning}
V.~Jeudy, R.~D. Pardo, W.~S. Torres, S.~Bustingorry, and A.~Kolton, ``Pinning
  of domain walls in thin ferromagnetic films,'' {\em Physical Review B},
  vol.~98, no.~5, p.~054406, 2018.

\bibitem{catarina2019magneto}
G.~Catarina, N.~Peres, and J.~Fern{\'a}ndez-Rossier, ``Magneto-optical {K}err
  effect in spin split two-dimensional massive {D}irac materials,'' {\em 2D
  Materials}, vol.~7, no.~2, p.~025011, 2020.

\bibitem{thiel2019probing}
L.~Thiel, Z.~Wang, M.~A. Tschudin, D.~Rohner, I.~Guti{\'e}rrez-Lezama,
  N.~Ubrig, M.~Gibertini, E.~Giannini, A.~F. Morpurgo, and P.~Maletinsky,
  ``Probing magnetism in 2{D} materials at the nanoscale with single-spin
  microscopy,'' {\em Science}, vol.~364, no.~6444, pp.~973--976, 2019.

\bibitem{matricardi1967electron}
V.~Matricardi, W.~Lehmann, N.~Kitamura, and J.~Silcox, ``Electron microscope
  observations of ferromagnetic domains in chromium tribromide,'' {\em Journal
  of Applied Physics}, vol.~38, no.~3, pp.~1297--1298, 1967.

\bibitem{virot2012theory}
F.~Virot, L.~Favre, R.~Hayn, and M.~Kuz'Min, ``Theory of magnetic domains in
  uniaxial thin films,'' {\em Journal of Physics D: Applied Physics}, vol.~45,
  no.~40, p.~405003, 2012.

\bibitem{kuhlow1975magnetic}
B.~Kuhlow and M.~Lambeck, ``Magnetic domain structures in {CrBr}$_3$,'' {\em
  Physica B+ C}, vol.~80, no.~1-4, pp.~365--373, 1975.

\end{thebibliography}


\begin{thebibliography}{12}%
\makeatletter
\providecommand \@ifxundefined [1]{%
 \@ifx{#1\undefined}
}%
\providecommand \@ifnum [1]{%
 \ifnum #1\expandafter \@firstoftwo
 \else \expandafter \@secondoftwo
 \fi
}%
\providecommand \@ifx [1]{%
 \ifx #1\expandafter \@firstoftwo
 \else \expandafter \@secondoftwo
 \fi
}%
\providecommand \natexlab [1]{#1}%
\providecommand \enquote  [1]{``#1''}%
\providecommand \bibnamefont  [1]{#1}%
\providecommand \bibfnamefont [1]{#1}%
\providecommand \citenamefont [1]{#1}%
\providecommand \href@noop [0]{\@secondoftwo}%
\providecommand \href [0]{\begingroup \@sanitize@url \@href}%
\providecommand \@href[1]{\@@startlink{#1}\@@href}%
\providecommand \@@href[1]{\endgroup#1\@@endlink}%
\providecommand \@sanitize@url [0]{\catcode `\\12\catcode `\$12\catcode
  `\&12\catcode `\#12\catcode `\^12\catcode `\_12\catcode `\%12\relax}%
\providecommand \@@startlink[1]{}%
\providecommand \@@endlink[0]{}%
\providecommand \url  [0]{\begingroup\@sanitize@url \@url }%
\providecommand \@url [1]{\endgroup\@href {#1}{\urlprefix }}%
\providecommand \urlprefix  [0]{URL }%
\providecommand \Eprint [0]{\href }%
\providecommand \doibase [0]{http://dx.doi.org/}%
\providecommand \selectlanguage [0]{\@gobble}%
\providecommand \bibinfo  [0]{\@secondoftwo}%
\providecommand \bibfield  [0]{\@secondoftwo}%
\providecommand \translation [1]{[#1]}%
\providecommand \BibitemOpen [0]{}%
\providecommand \bibitemStop [0]{}%
\providecommand \bibitemNoStop [0]{.\EOS\space}%
\providecommand \EOS [0]{\spacefactor3000\relax}%
\providecommand \BibitemShut  [1]{\csname bibitem#1\endcsname}%
\let\auto@bib@innerbib\@empty
\bibitem [{\citenamefont {Giannozzi}\ \emph {et~al.}(2009)\citenamefont
  {Giannozzi}, \citenamefont {Baroni}, \citenamefont {Bonini}, \citenamefont
  {Calandra}, \citenamefont {Car}, \citenamefont {Cavazzoni}, \citenamefont
  {Ceresoli}, \citenamefont {Chiarotti}, \citenamefont {Cococcioni},
  \citenamefont {Dabo} \emph {et~al.}}]{Giannozzi2009}%
  \BibitemOpen
  \bibfield  {author} {\bibinfo {author} {\bibfnamefont {P.}~\bibnamefont
  {Giannozzi}}, \bibinfo {author} {\bibfnamefont {S.}~\bibnamefont {Baroni}},
  \bibinfo {author} {\bibfnamefont {N.}~\bibnamefont {Bonini}}, \bibinfo
  {author} {\bibfnamefont {M.}~\bibnamefont {Calandra}}, \bibinfo {author}
  {\bibfnamefont {R.}~\bibnamefont {Car}}, \bibinfo {author} {\bibfnamefont
  {C.}~\bibnamefont {Cavazzoni}}, \bibinfo {author} {\bibfnamefont
  {D.}~\bibnamefont {Ceresoli}}, \bibinfo {author} {\bibfnamefont {G.~L.}\
  \bibnamefont {Chiarotti}}, \bibinfo {author} {\bibfnamefont {M.}~\bibnamefont
  {Cococcioni}}, \bibinfo {author} {\bibfnamefont {I.}~\bibnamefont {Dabo}},
  \emph {et~al.},\ }\bibfield  {title} {\emph {\bibinfo {title} {Quantum
  espresso: a modular and open-source software project for quantum simulations
  of materials},\ }}\href@noop {} {\bibfield  {journal} {\bibinfo  {journal}
  {Journal of physics: Condensed matter}\ }\textbf {\bibinfo {volume} {21}},\
  \bibinfo {pages} {395502} (\bibinfo {year} {2009})}\BibitemShut {NoStop}%
\bibitem [{\citenamefont {Wu}\ \emph {et~al.}(2019)\citenamefont {Wu},
  \citenamefont {Li}, \citenamefont {Cao},\ and\ \citenamefont
  {Louie}}]{Wu2019}%
  \BibitemOpen
  \bibfield  {author} {\bibinfo {author} {\bibfnamefont {M.}~\bibnamefont
  {Wu}}, \bibinfo {author} {\bibfnamefont {Z.}~\bibnamefont {Li}}, \bibinfo
  {author} {\bibfnamefont {T.}~\bibnamefont {Cao}}, \ and\ \bibinfo {author}
  {\bibfnamefont {S.~G.}\ \bibnamefont {Louie}},\ }\bibfield  {title} {\emph
  {\bibinfo {title} {{Physical origin of giant excitonic and magneto-optical
  responses in two-dimensional ferromagnetic insulators}},\ }}\href {\doibase
  10.1038/s41467-019-10325-7} {\bibfield  {journal} {\bibinfo  {journal}
  {Nature Communications}\ }\textbf {\bibinfo {volume} {10}},\ \bibinfo {pages}
  {2371} (\bibinfo {year} {2019})}\BibitemShut {NoStop}%
\bibitem [{\citenamefont {Hamann}(2013)}]{Hamann2013}%
  \BibitemOpen
  \bibfield  {author} {\bibinfo {author} {\bibfnamefont {D.~R.}\ \bibnamefont
  {Hamann}},\ }\bibfield  {title} {\emph {\bibinfo {title} {{Optimized
  norm-conserving Vanderbilt pseudopotentials}},\ }}\href {\doibase
  10.1103/PhysRevB.88.085117} {\bibfield  {journal} {\bibinfo  {journal}
  {Physical Review B}\ }\textbf {\bibinfo {volume} {88}},\ \bibinfo {pages}
  {085117} (\bibinfo {year} {2013})}\BibitemShut {NoStop}%
\bibitem [{\citenamefont {van Setten}\ \emph {et~al.}(2018)\citenamefont {van
  Setten}, \citenamefont {Giantomassi}, \citenamefont {Bousquet}, \citenamefont
  {Verstraete}, \citenamefont {Hamann}, \citenamefont {Gonze},\ and\
  \citenamefont {Rignanese}}]{VanSetten2018}%
  \BibitemOpen
  \bibfield  {author} {\bibinfo {author} {\bibfnamefont {M.}~\bibnamefont {van
  Setten}}, \bibinfo {author} {\bibfnamefont {M.}~\bibnamefont {Giantomassi}},
  \bibinfo {author} {\bibfnamefont {E.}~\bibnamefont {Bousquet}}, \bibinfo
  {author} {\bibfnamefont {M.}~\bibnamefont {Verstraete}}, \bibinfo {author}
  {\bibfnamefont {D.}~\bibnamefont {Hamann}}, \bibinfo {author} {\bibfnamefont
  {X.}~\bibnamefont {Gonze}}, \ and\ \bibinfo {author} {\bibfnamefont {G.-M.}\
  \bibnamefont {Rignanese}},\ }\bibfield  {title} {\emph {\bibinfo {title}
  {{The PseudoDojo: Training and grading a 85 element optimized norm-conserving
  pseudopotential table}},\ }}\href {\doibase 10.1016/J.CPC.2018.01.012}
  {\bibfield  {journal} {\bibinfo  {journal} {Computer Physics Communications}\
  }\textbf {\bibinfo {volume} {226}},\ \bibinfo {pages} {39} (\bibinfo {year}
  {2018})}\BibitemShut {NoStop}%
\bibitem [{\citenamefont {Popescu}\ and\ \citenamefont
  {Zunger}(2012)}]{Popescu2012}%
  \BibitemOpen
  \bibfield  {author} {\bibinfo {author} {\bibfnamefont {V.}~\bibnamefont
  {Popescu}}\ and\ \bibinfo {author} {\bibfnamefont {A.}~\bibnamefont
  {Zunger}},\ }\bibfield  {title} {\emph {\bibinfo {title} {Extracting {$E$}
  versus $\vec{k}$ effective band structure from supercell calculations on
  alloys and impurities},\ }}\href {\doibase 10.1103/PhysRevB.85.085201}
  {\bibfield  {journal} {\bibinfo  {journal} {Phys. Rev. B}\ }\textbf {\bibinfo
  {volume} {85}},\ \bibinfo {pages} {085201} (\bibinfo {year}
  {2012})}\BibitemShut {NoStop}%
\bibitem [{\citenamefont {Molina-S\'anchez}\ \emph {et~al.}(2015)\citenamefont
  {Molina-S\'anchez}, \citenamefont {Hummer},\ and\ \citenamefont
  {Wirtz}}]{Molina-Sanchez2015}%
  \BibitemOpen
  \bibfield  {author} {\bibinfo {author} {\bibfnamefont {A.}~\bibnamefont
  {Molina-S\'anchez}}, \bibinfo {author} {\bibfnamefont {K.}~\bibnamefont
  {Hummer}}, \ and\ \bibinfo {author} {\bibfnamefont {L.}~\bibnamefont
  {Wirtz}},\ }\bibfield  {title} {\emph {\bibinfo {title} {Vibrational and
  optical properties of {MoS}2: {From} monolayer to bulk},\ }}\href {\doibase
  10.1016/j.surfrep.2015.10.001} {\bibfield  {journal} {\bibinfo  {journal}
  {Surface Science Reports}\ }\textbf {\bibinfo {volume} {70}},\ \bibinfo
  {pages} {554} (\bibinfo {year} {2015})}\BibitemShut {NoStop}%
\bibitem [{\citenamefont {Ghazaryan}\ \emph {et~al.}(2018)\citenamefont
  {Ghazaryan}, \citenamefont {Greenaway}, \citenamefont {Wang}, \citenamefont
  {Guarochico-Moreira}, \citenamefont {Vera-Marun}, \citenamefont {Yin},
  \citenamefont {Liao}, \citenamefont {Morozov}, \citenamefont {Kristanovski},
  \citenamefont {Lichtenstein} \emph {et~al.}}]{ghazaryan2018magnon}%
  \BibitemOpen
  \bibfield  {author} {\bibinfo {author} {\bibfnamefont {D.}~\bibnamefont
  {Ghazaryan}}, \bibinfo {author} {\bibfnamefont {M.~T.}\ \bibnamefont
  {Greenaway}}, \bibinfo {author} {\bibfnamefont {Z.}~\bibnamefont {Wang}},
  \bibinfo {author} {\bibfnamefont {V.~H.}\ \bibnamefont {Guarochico-Moreira}},
  \bibinfo {author} {\bibfnamefont {I.~J.}\ \bibnamefont {Vera-Marun}},
  \bibinfo {author} {\bibfnamefont {J.}~\bibnamefont {Yin}}, \bibinfo {author}
  {\bibfnamefont {Y.}~\bibnamefont {Liao}}, \bibinfo {author} {\bibfnamefont
  {S.~V.}\ \bibnamefont {Morozov}}, \bibinfo {author} {\bibfnamefont
  {O.}~\bibnamefont {Kristanovski}}, \bibinfo {author} {\bibfnamefont {A.~I.}\
  \bibnamefont {Lichtenstein}},  \emph {et~al.},\ }\bibfield  {title} {\emph
  {\bibinfo {title} {Magnon-assisted tunnelling in van der {Waals}
  heterostructures based on {CrBr$_3$}},\ }}\href@noop {} {\bibfield  {journal}
  {\bibinfo  {journal} {Nature Electronics}\ }\textbf {\bibinfo {volume} {1}},\
  \bibinfo {pages} {344} (\bibinfo {year} {2018})}\BibitemShut {NoStop}%
\bibitem [{\citenamefont {Dufferwiel}\ \emph {et~al.}(2017)\citenamefont
  {Dufferwiel}, \citenamefont {Lyons}, \citenamefont {Solnyshkov},
  \citenamefont {Trichet}, \citenamefont {Withers}, \citenamefont {Schwarz},
  \citenamefont {Malpuech}, \citenamefont {Smith}, \citenamefont {Novoselov},
  \citenamefont {Skolnick}, \citenamefont {Krizhanovskii},\ and\ \citenamefont
  {Tartakovskii}}]{dufferwiel2017valley}%
  \BibitemOpen
  \bibfield  {author} {\bibinfo {author} {\bibfnamefont {S.}~\bibnamefont
  {Dufferwiel}}, \bibinfo {author} {\bibfnamefont {T.~P.}\ \bibnamefont
  {Lyons}}, \bibinfo {author} {\bibfnamefont {D.~D.}\ \bibnamefont
  {Solnyshkov}}, \bibinfo {author} {\bibfnamefont {A.~A.~P.}\ \bibnamefont
  {Trichet}}, \bibinfo {author} {\bibfnamefont {F.}~\bibnamefont {Withers}},
  \bibinfo {author} {\bibfnamefont {S.}~\bibnamefont {Schwarz}}, \bibinfo
  {author} {\bibfnamefont {G.}~\bibnamefont {Malpuech}}, \bibinfo {author}
  {\bibfnamefont {J.~M.}\ \bibnamefont {Smith}}, \bibinfo {author}
  {\bibfnamefont {K.~S.}\ \bibnamefont {Novoselov}}, \bibinfo {author}
  {\bibfnamefont {M.~S.}\ \bibnamefont {Skolnick}}, \bibinfo {author}
  {\bibfnamefont {D.~N.}\ \bibnamefont {Krizhanovskii}}, \ and\ \bibinfo
  {author} {\bibfnamefont {A.~I.}\ \bibnamefont {Tartakovskii}},\ }\bibfield
  {title} {\emph {\bibinfo {title} {Valley-addressable polaritons in atomically
  thin semiconductors},\ }}\href@noop {} {\bibfield  {journal} {\bibinfo
  {journal} {Nature Photonics}\ }\textbf {\bibinfo {volume} {11}},\ \bibinfo
  {pages} {497} (\bibinfo {year} {2017})}\BibitemShut {NoStop}%
\bibitem [{\citenamefont {Wang}\ \emph {et~al.}(2015)\citenamefont {Wang},
  \citenamefont {Palleau}, \citenamefont {Amand}, \citenamefont {Tongay},
  \citenamefont {Marie},\ and\ \citenamefont
  {Urbaszek}}]{wang2015polarization}%
  \BibitemOpen
  \bibfield  {author} {\bibinfo {author} {\bibfnamefont {G.}~\bibnamefont
  {Wang}}, \bibinfo {author} {\bibfnamefont {E.}~\bibnamefont {Palleau}},
  \bibinfo {author} {\bibfnamefont {T.}~\bibnamefont {Amand}}, \bibinfo
  {author} {\bibfnamefont {S.}~\bibnamefont {Tongay}}, \bibinfo {author}
  {\bibfnamefont {X.}~\bibnamefont {Marie}}, \ and\ \bibinfo {author}
  {\bibfnamefont {B.}~\bibnamefont {Urbaszek}},\ }\bibfield  {title} {\emph
  {\bibinfo {title} {Polarization and time-resolved photoluminescence
  spectroscopy of excitons in {MoSe}$_2$ monolayers},\ }}\href@noop {}
  {\bibfield  {journal} {\bibinfo  {journal} {Applied Physics Letters}\
  }\textbf {\bibinfo {volume} {106}},\ \bibinfo {pages} {112101} (\bibinfo
  {year} {2015})}\BibitemShut {NoStop}%
\bibitem [{\citenamefont {Glazov}\ \emph {et~al.}(2014)\citenamefont {Glazov},
  \citenamefont {Amand}, \citenamefont {Marie}, \citenamefont {Lagarde},
  \citenamefont {Bouet},\ and\ \citenamefont {Urbaszek}}]{glazov2014exciton}%
  \BibitemOpen
  \bibfield  {author} {\bibinfo {author} {\bibfnamefont {M.}~\bibnamefont
  {Glazov}}, \bibinfo {author} {\bibfnamefont {T.}~\bibnamefont {Amand}},
  \bibinfo {author} {\bibfnamefont {X.}~\bibnamefont {Marie}}, \bibinfo
  {author} {\bibfnamefont {D.}~\bibnamefont {Lagarde}}, \bibinfo {author}
  {\bibfnamefont {L.}~\bibnamefont {Bouet}}, \ and\ \bibinfo {author}
  {\bibfnamefont {B.}~\bibnamefont {Urbaszek}},\ }\bibfield  {title} {\emph
  {\bibinfo {title} {Exciton fine structure and spin decoherence in monolayers
  of transition metal dichalcogenides},\ }}\href@noop {} {\bibfield  {journal}
  {\bibinfo  {journal} {Physical Review B}\ }\textbf {\bibinfo {volume} {89}},\
  \bibinfo {pages} {201302} (\bibinfo {year} {2014})}\BibitemShut {NoStop}%
\bibitem [{\citenamefont {Maialle}\ \emph {et~al.}(1993)\citenamefont
  {Maialle}, \citenamefont {e~Silva},\ and\ \citenamefont
  {Sham}}]{maialle1993exciton}%
  \BibitemOpen
  \bibfield  {author} {\bibinfo {author} {\bibfnamefont {M.~Z.}\ \bibnamefont
  {Maialle}}, \bibinfo {author} {\bibfnamefont {E.~A. d.~A.}\ \bibnamefont
  {e~Silva}}, \ and\ \bibinfo {author} {\bibfnamefont {L.~J.}\ \bibnamefont
  {Sham}},\ }\bibfield  {title} {\emph {\bibinfo {title} {Exciton spin dynamics
  in quantum wells},\ }}\href@noop {} {\bibfield  {journal} {\bibinfo
  {journal} {Physical Review B}\ }\textbf {\bibinfo {volume} {47}},\ \bibinfo
  {pages} {15776} (\bibinfo {year} {1993})}\BibitemShut {NoStop}%
\bibitem [{\citenamefont {Li}\ \emph {et~al.}(2014)\citenamefont {Li},
  \citenamefont {Ludwig}, \citenamefont {Low}, \citenamefont {Chernikov},
  \citenamefont {Cui}, \citenamefont {Arefe}, \citenamefont {Kim},
  \citenamefont {Van Der~Zande}, \citenamefont {Rigosi}, \citenamefont {Hill}
  \emph {et~al.}}]{li2014valley}%
  \BibitemOpen
  \bibfield  {author} {\bibinfo {author} {\bibfnamefont {Y.}~\bibnamefont
  {Li}}, \bibinfo {author} {\bibfnamefont {J.}~\bibnamefont {Ludwig}}, \bibinfo
  {author} {\bibfnamefont {T.}~\bibnamefont {Low}}, \bibinfo {author}
  {\bibfnamefont {A.}~\bibnamefont {Chernikov}}, \bibinfo {author}
  {\bibfnamefont {X.}~\bibnamefont {Cui}}, \bibinfo {author} {\bibfnamefont
  {G.}~\bibnamefont {Arefe}}, \bibinfo {author} {\bibfnamefont {Y.~D.}\
  \bibnamefont {Kim}}, \bibinfo {author} {\bibfnamefont {A.~M.}\ \bibnamefont
  {Van Der~Zande}}, \bibinfo {author} {\bibfnamefont {A.}~\bibnamefont
  {Rigosi}}, \bibinfo {author} {\bibfnamefont {H.~M.}\ \bibnamefont {Hill}},
  \emph {et~al.},\ }\bibfield  {title} {\emph {\bibinfo {title} {Valley
  splitting and polarization by the {Z}eeman effect in monolayer {MoSe}$_2$},\
  }}\href@noop {} {\bibfield  {journal} {\bibinfo  {journal} {Physical Review
  Letters}\ }\textbf {\bibinfo {volume} {113}},\ \bibinfo {pages} {266804}
  (\bibinfo {year} {2014})}\BibitemShut {NoStop}%
\end{thebibliography}%

	

	
\end{document}


\title{Supplementary Information for: Interplay between spin proximity effect and charge-dependent exciton dynamics in MoSe$_2$ / CrBr$_3$ van der Waals heterostructures}

\author{T. P. Lyons}
\email{t.lyons@sheffield.ac.uk}
\author{D. Gillard}
\affiliation{Department of Physics and Astronomy, The University of Sheffield, Sheffield S3 7RH, UK}
\author{A. Molina-S\'anchez}
\affiliation{QuantaLab, International Iberian Nanotechnology Laboratory, 4715-330 Braga, Portugal}
\author{A. Misra}
\affiliation{School of Physics and Astronomy, The University of Manchester, Manchester M13 9PL, UK}
\affiliation{Department of Physics, Indian Institute of Technology Madras (IIT Madras), Chennai, India}
\author{F. Withers}
\affiliation{College of Engineering, Mathematics and Physical Sciences, University of Exeter, Exeter, EX4 4QF, UK}
\author{P. S. Keatley}
\affiliation{College of Engineering, Mathematics and Physical Sciences, University of Exeter, Exeter, EX4 4QF, UK}
\author{A. Kozikov}
\affiliation{School of Physics and Astronomy, The University of Manchester, Manchester M13 9PL, UK}
\author{T. Taniguchi}
\affiliation{National Institute for Materials Science, Tsukuba, Ibaraki 305-0044, Japan}
\author{K. Watanabe}
\affiliation{National Institute for Materials Science, Tsukuba, Ibaraki 305-0044, Japan}
\author{K. S. Novoselov}
\affiliation{School of Physics and Astronomy, The University of Manchester, Manchester M13 9PL, UK}
\affiliation{Centre for Advanced 2D Materials, National University of Singapore, 117546 Singapore}
\affiliation{Chongqing 2D Materials Institute, Liangjiang New Area, Chongqing, 400714, China}
\author{J. Fern\'andez-Rossier}
\affiliation{QuantaLab, International Iberian Nanotechnology Laboratory, 4715-330 Braga, Portugal}
\author{A. I. Tartakovskii}
\email{a.tartakovskii@sheffield.ac.uk}
\affiliation{Department of Physics and Astronomy, The University of Sheffield, Sheffield S3 7RH, UK}
\maketitle

\clearpage

\subsection{\textbf{Supplementary Note 1: Details of the DFT calculations}}

The \textit{ab initio} calculations of the electronic structure of the heterostructure MoSe$_2$/CrBr$_3$ have been performed using density-functional theory (DFT) at the level of the local density approximation (LDA), as implemented in Quantum Espresso package \cite{Giannozzi2009}. In addition we have applied the on-site Hubbard correction with values U=1.5 eV and Hund’s exchange interaction J=0.5 eV \cite{Wu2019}. We have included spin-orbit interaction with spinorial wave functions, using norm-conserving full relativistic pseudopotentials. The pseudopotentials of Cr and Mo include semi-core valence electrons and have been generated with ONCVPSP and PSEUDOJO \cite{Hamann2013,VanSetten2018}. The electronic density converges with an energy cutoff of 80 Ry and a $\bf{k}$-grid of $12\times 12\times 1$. We use a slab model with a 17 \AA~ vacuum thickness to avoid interactions between periodic images.

In order to make the interpretation of the band structure easier we have followed the unfolding procedure formulated by Popescu and Zunger \cite{Popescu2012}. This method maps the energy eigenvalues obtained in supercell calculations into an effective band structure (EBS). The reference band structure is that of monolayer MoSe$_2$.	

We have performed two sets of \textit{ab initio} calculations: (i) imposing the experimental lattice constant of MoSe$_2$ and relaxing the atomic positions of the whole heterostructure, and (ii) relaxing the lattice parameter of the supercell. In this way we assure that no artifact from the simulation alters the conclusions. Figure 1a-b shows the EBS material-projected and spin-projected, respectively, for the case of MoSe$_2$ experimental lattice constant. We find the strongest hybridization of the EBS at the intermediate Q point in the conduction band. Moreover, the conduction band is immersed within the $t_{2g}$ bands of CrBr$_3$. Figure 1c shows the detail of the conduction band at the $K$ and $K$' points of the EBS, which reveals a valley splitting of 2.2 meV. In free-standing MoSe$_2$ and without magnetic field, conduction band states at $K$ and $K$' have the same energy. Our calculations indicate no valley splitting in the valence band. 

\begin{figure*}[h]
\begin{center}
\includegraphics[scale=0.65]{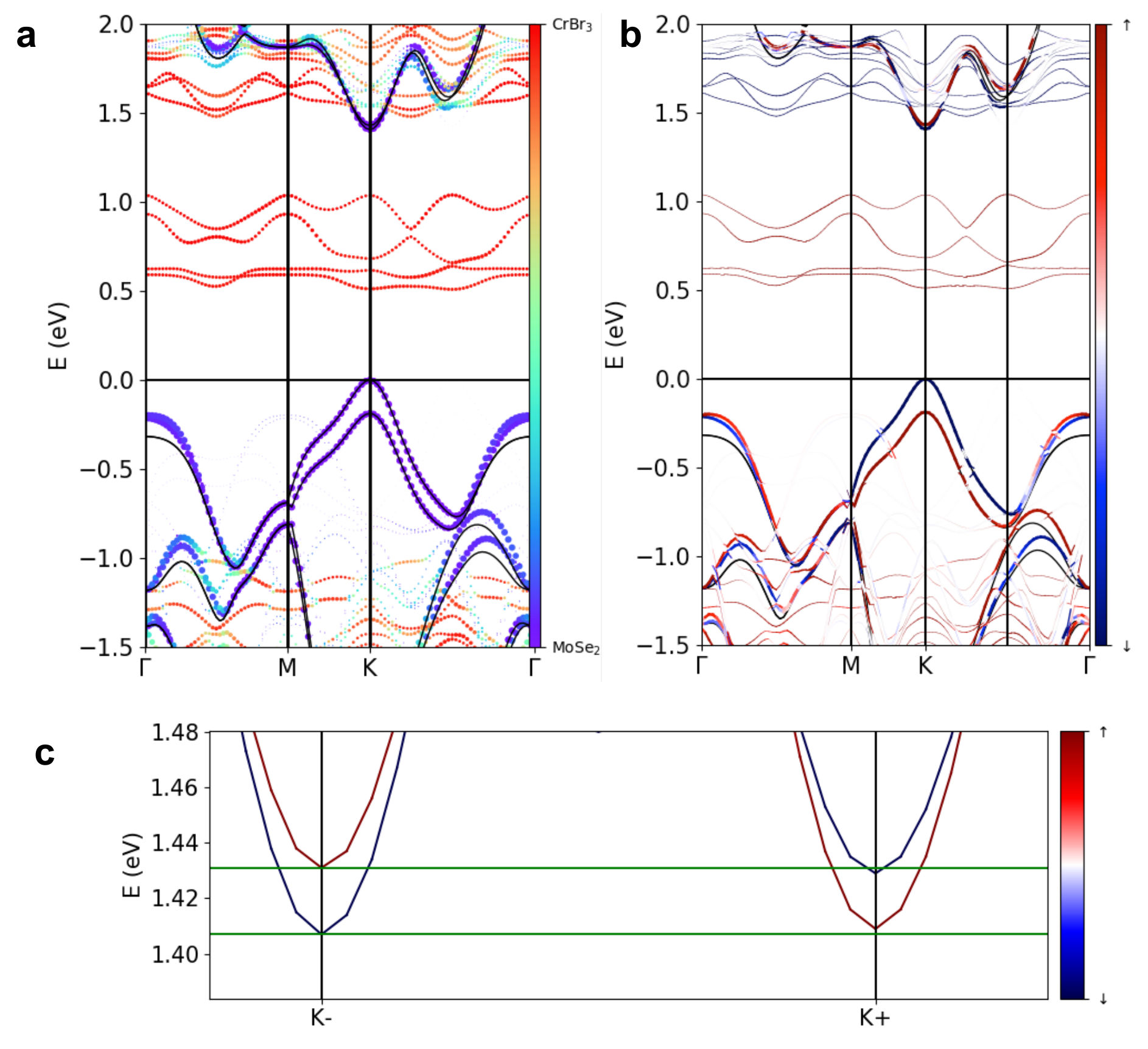}
\end{center}
\caption{Effective band structure of MoSe$_2$/CrBr$_3$ with MoSe$_2$ experimental lattice constant. (a) Material-projected EBS, (b) spin-projected EBS and (c) MoSe$_2$ conduction bands around $K$ and $K'$, revealing a valley splitting of 2.2 meV.}
\label{ebs-exp}
\end{figure*}

\clearpage

Figure 2 shows the relevant wavefunctions at important points in the EBS at the K-point. There is evidence of interlayer band hybridization in the MoSe$_2$ conduction band, corresponding to box 5 in Figure 2. The orbital content of the MoSe$_2$ valence band is localized primarily in the Molybdenum plane (box 1), and so experiences a far weaker spin proximity effect as compared to the MoSe$_2$ conduction band, in which the orbital spread is more prominent out of the plane (boxes 3 and 4). This explains the breaking of valley degeneracy in the conduction band (Figure 1c), absent for the valence band.  

\begin{figure*}[h]
\begin{center}
\includegraphics[scale=0.55]{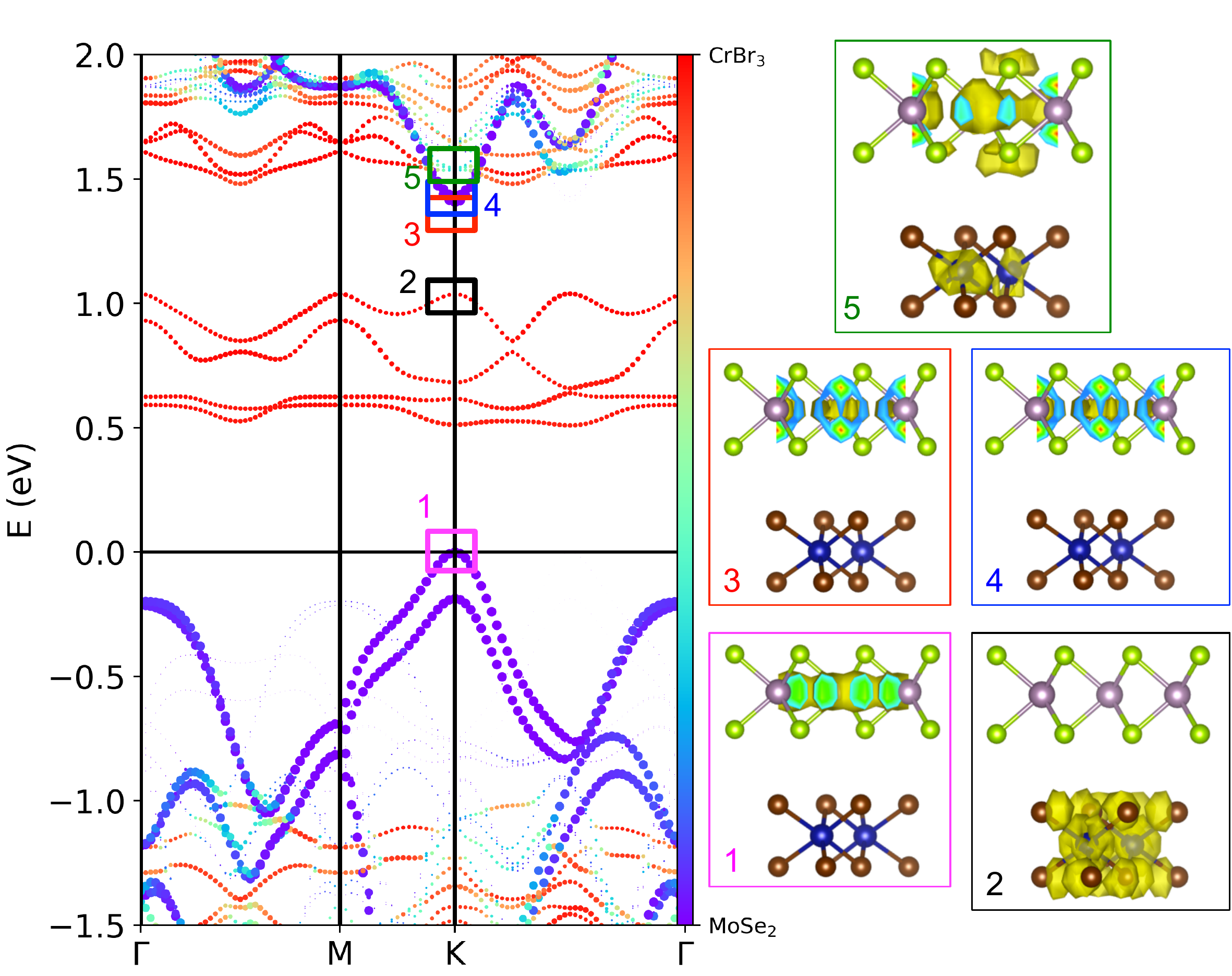}
\end{center}
\caption{Band structure of MoSe$_2$ / CrBr$_3$ (as in Fig. 1e of Main Text) together with the wave function density of the following band states: 1. Top of valence band (MoSe$_2$); 2. Conduction band $e_g$ of CrBr$_3$; 3 and 4. Conduction band states of MoSe$_2$ spin up and down; 5. Conduction band state MoSe$_2$-CrBr$_3$. }
\label{ebs-exp}
\end{figure*}

\clearpage

We have also evaluated the EBS for an optimized heterostructure, as shown in Figure 3. In this case we find an indirect bandgap for MoSe$_2$, and a stronger hybridization of the conduction band at the Q-point. This is a well-known consequence of the dependency of the bandgap directness on the lattice parameter in transition metal dichalcogenides \cite{Molina-Sanchez2015}. In this case a conduction band valley splitting is also expected, as shown in Figure 3c.

\begin{figure*}[h]
\begin{center}
\includegraphics[scale=0.65]{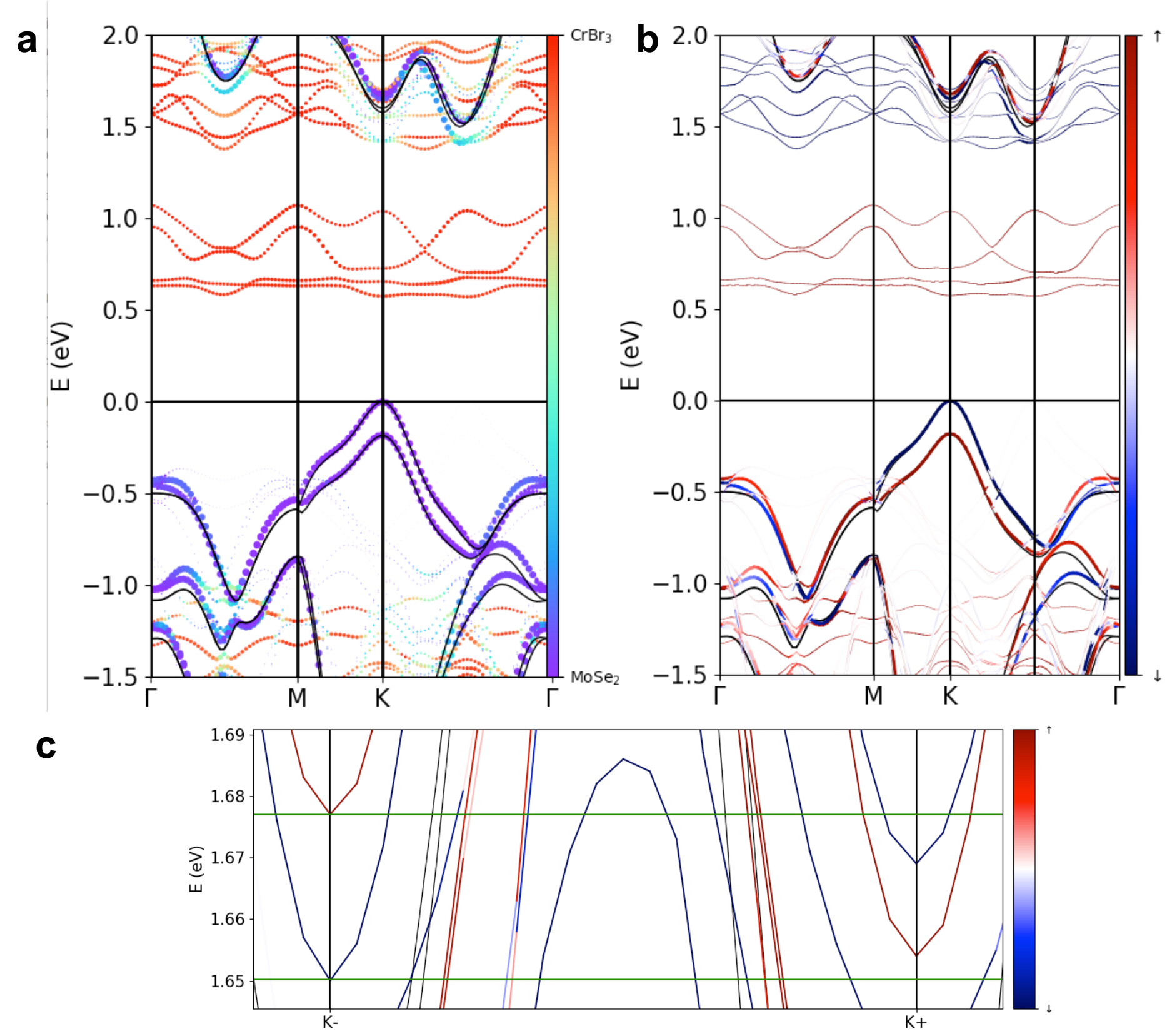}
\end{center}
\caption{Effective band structure of MoSe$_2$/CrBr$_3$ with optimized lattice parameters. (a) Material-projected EBS, (b) spin-projected EBS and (c) conduction bands around $K$ and $K'$.}
\label{ebs-opt}
\end{figure*}

\clearpage

\subsection{\textbf{Supplementary Note 2: Band structure of monolayer-MoSe$_2$ on bilayer-CrBr$_3$}}

We have confirmed the lack of significant contribution to the proximity effects from underlying CrBr$_3$ layers by comparing the electronic structure of monolayer MoSe$_2$ on top of single-layer and double-layer CrBr$_3$. Essentially we find that adding more layers results in the increasing of states from CrBr$_3$.

\begin{figure*}[h]
\begin{center}
\includegraphics[scale=0.6]{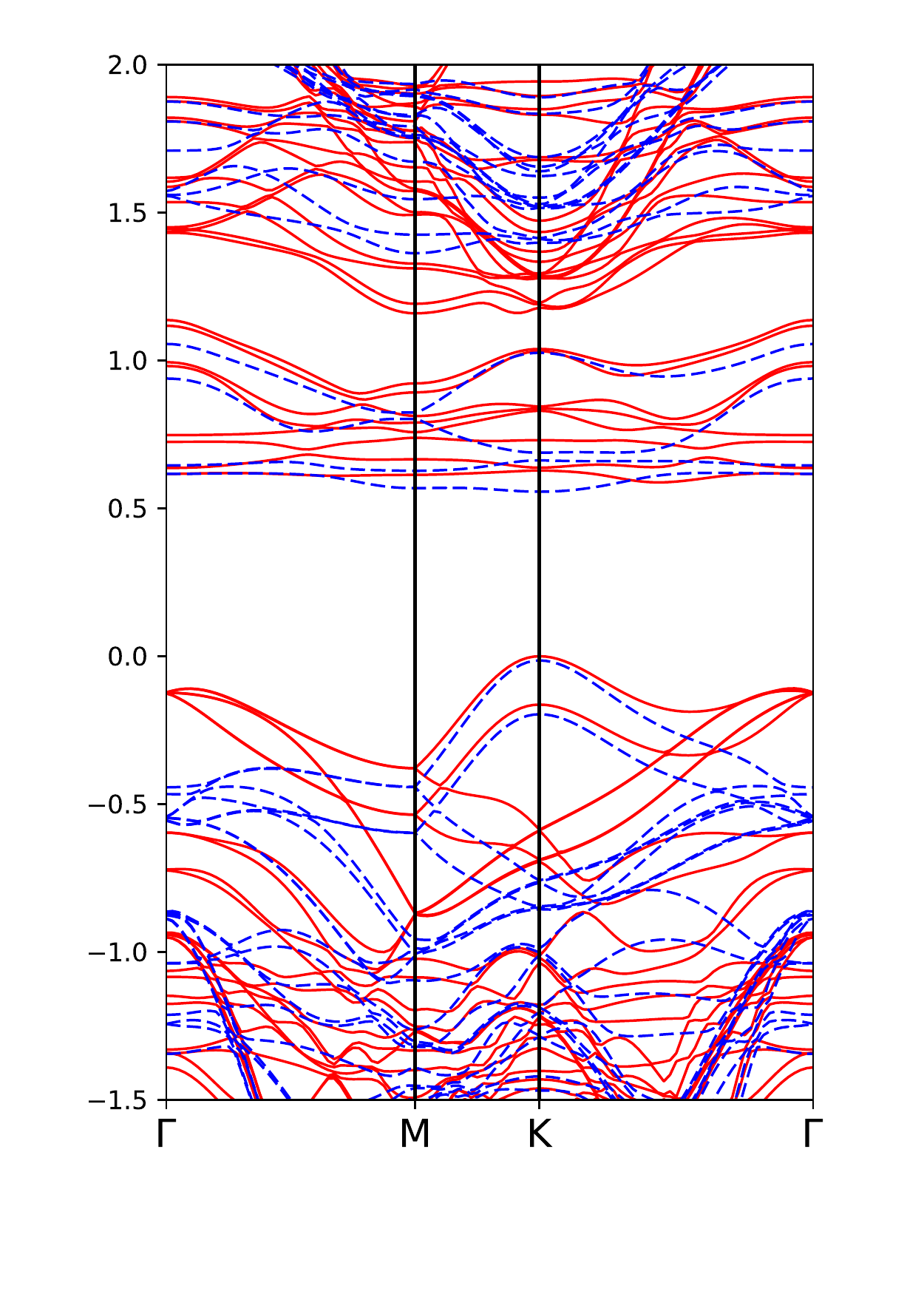}
\end{center}
\caption{Comparison of effective band structures for the cases of MoSe$_2$ on top of monolayer CrBr$_3$ (blue dashed lines) or MoSe$_2$ on top of bilayer CrBr$_3$ (red lines). The thicker CrBr$_3$ adds more bands to the EBS as a whole, and the absolute energies of some of the bands shift, but the $e_g$ bands remain at several hundred millielectron Volts below the $t_{2g}$ bands, and so the spin-dependent interlayer tunnelling discussed in the Main Text is not expected to change substantially.}
\label{ebs-opt}
\end{figure*}

\clearpage

\subsection{Supplementary Note 3: Measurements on additional samples}

We have fabricated and measured the DOCP in photoluminescence of 2 additional MoSe$_2$ / CrBr$_3$ samples. The results are presented below. As can be seen, both samples reproduce the core finding from the sample presented in the Main Text, that is, exciton insensitivity to the proximity effects, alongside trion DOCP switching (evidenced by the sharp discontinuity, present only in the trion state, in the colourmaps below) arising from spin-dependent interlayer charge transfer.

We note some minor differences attributable to sample variation. This is to be expected, especially considering that these additional samples were not fabricated using the same bulk CrBr$_3$ crystal as Sample 1 presented in the Main Text. For instance, the exact shape of the trion DOCP vs B is not identical to Sample 1. This reflects a difference in ferromagnetic domain dynamics. As discussed in the Main Text, the domain formation patterns and sizes depend heavily on a host of factors which will inevitably vary from one flake to the next. We also note that CrBr$_3$ is very unstable and sensitive to degradation. We cannot be sure how many layers within each flake are still magnetically active, and how that may influence the domain dynamics of the flake as a whole.

The other noticeable difference is that the trion DOCP in Sample 3 does not cross DOCP = 0. This is a linear offset in DOCP which does not influence the hysteresis behaviour. We expect that it arises from a portion of the CrBr$_3$ flake which is not responsive to B-field, as the magnetization may be pinned by disorder or degradation. In all 3 samples, only the trion state is sensitive to CrBr$_3$ magnetization fluctuations, while the exciton displays only a shallow gradient owing to the conventional valley Zeeman effect, rather than any proximity interactions.

\begin{figure*}[h]
	\begin{center}
		\includegraphics[width=15cm]{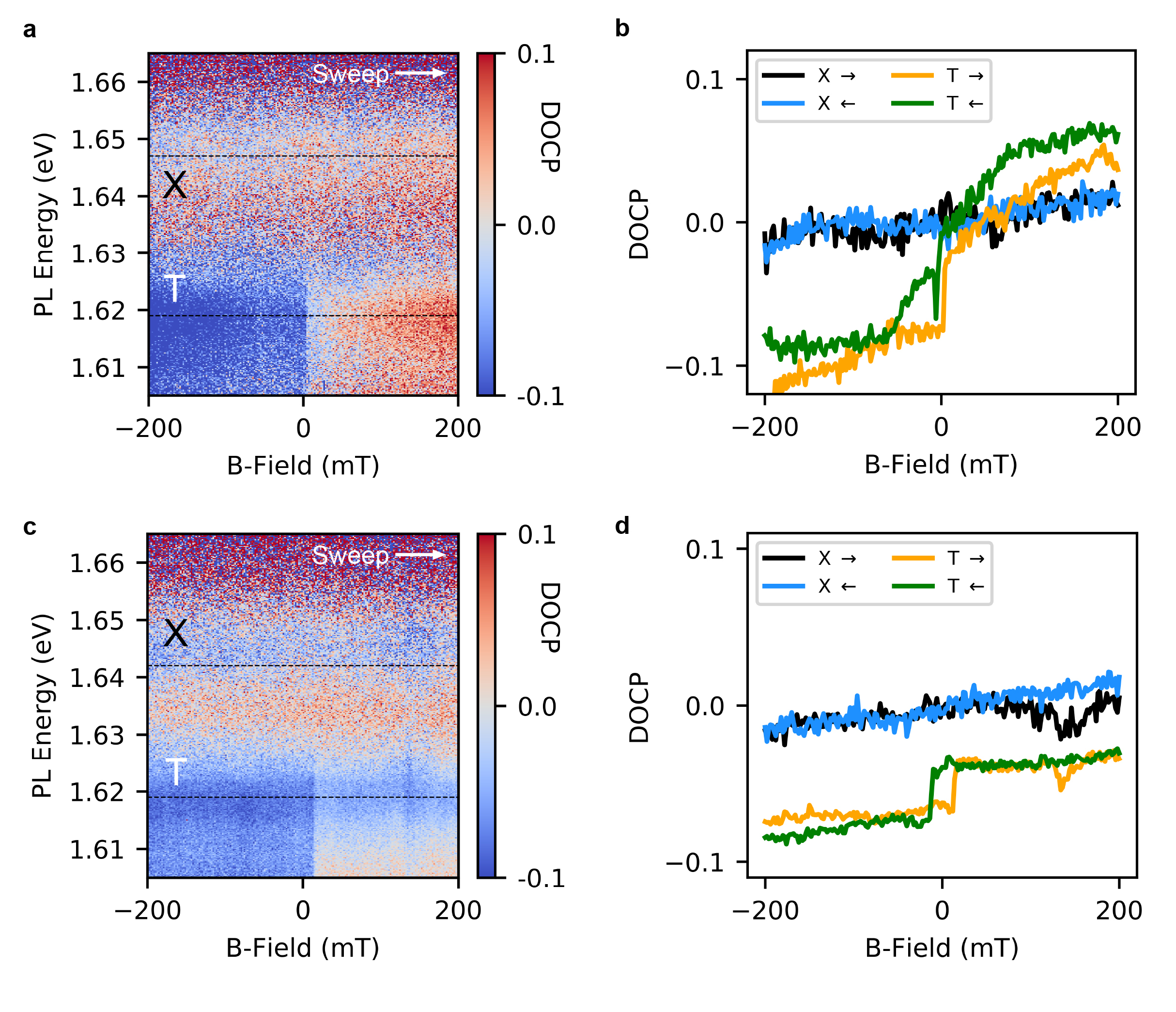}
	\end{center}
	\caption{Data from additional samples. (a,c) Degree of circular polarization (DOCP) from Sample 2 (a) and Sample 3 (c) in the forward sweep direction. (b,d) DOCP of exciton and trion in Sample 2 (b) and Sample 3 (d) calculated using integrated intensities, in both B-Field sweep directions. }
	\label{extra_samples}
\end{figure*}

\clearpage

\subsection{Supplementary Note 4: Temperature dependence of trion DOCP}

We heat the sample from the base temperature of 4.2 K up to 60 K, under a constant applied magnetic field of $B=+200$ mT, in order to saturate the CrBr$_3$ magnetization. The polarization degree is observed to decrease with increasing temperature, and to tend towards zero above the reported Curie temperature of CrBr$_3$ of $\sim 37$ K \cite{ghazaryan2018magnon}. We note that the CrBr$_3$ becomes paramagnetic above the Curie temperature, and so some spin polarization of the electronic bands is likely to persist above $\sim 37$ K owing to the influence of the external saturation field. Therefore, no sharp drop in DOCP is necessarily expected exactly at the Curie point.

\begin{figure*}[h]
\begin{center}
\includegraphics[scale=1]{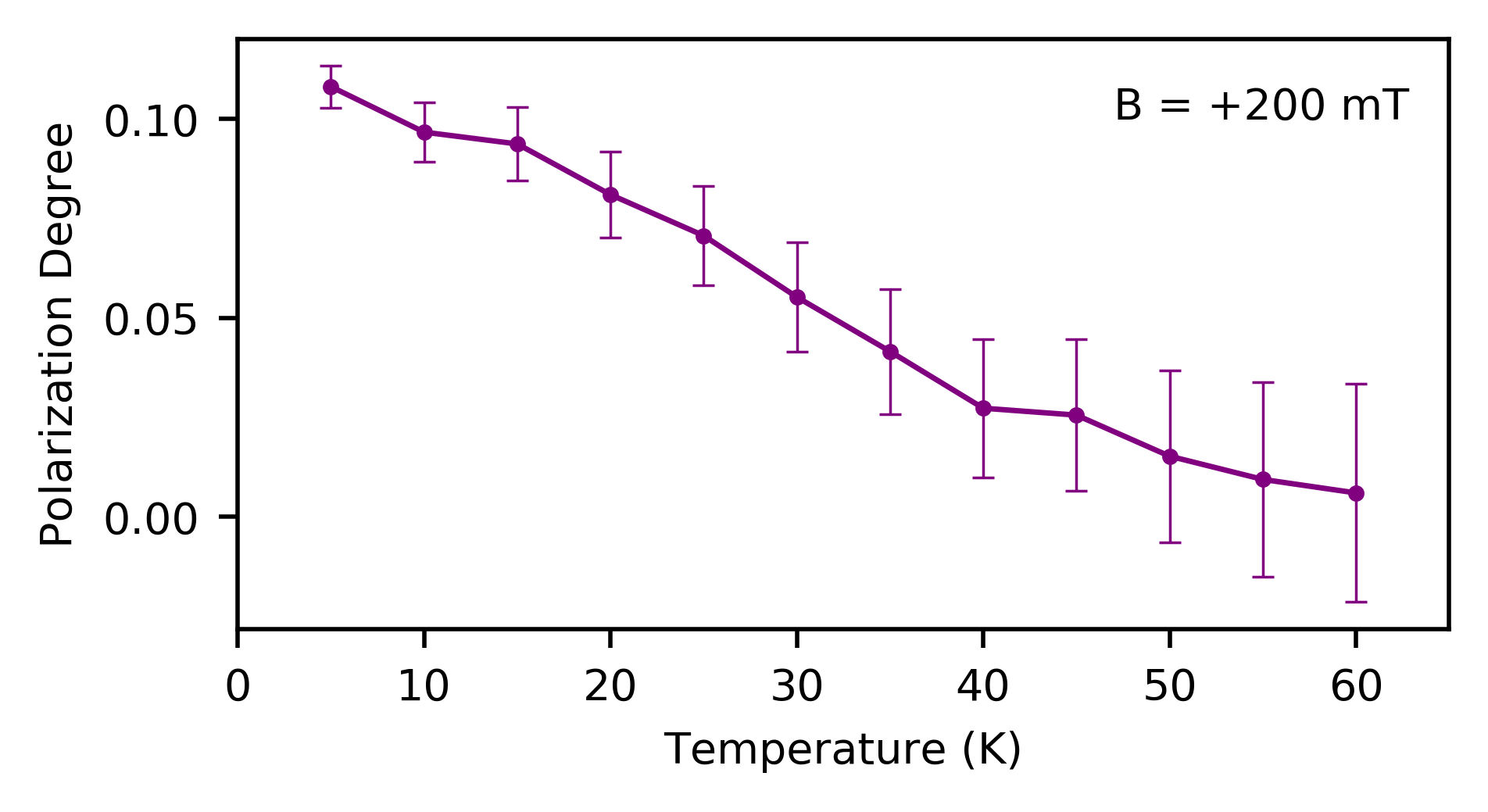}
\end{center}
\caption{Trion polarization degree as a function of sample temperature, while heated
under a fixed external B-field of strength 200 mT, corresponding to saturation of the CrBr$_3$
magnetization parallel to B. The error in the measurement increases with temperature as the PL signal intensity decreases. }
\label{temp-dep}
\end{figure*}

\clearpage

\subsection{Supplementary Note 5: Insensitivity of results to polarization state of laser}

The laser remains in $\sigma^+$ circular polarization for all data presented in the Main Text. Nominally, this polarization state addresses only the K valley of MoSe$_2$, rather than the K' valley, which in principle may introduce a finite valley polarization of photogenerated carriers, excitons or trions, leading to a non-zero circular polarization degree in the eventual photoluminescence. However, the lack of significant retention of non-resonantly optically injected valley polarization in photoluminescence from monolayer MoSe$_2$ is well known \cite{dufferwiel2017valley,wang2015polarization}, most commonly attributed to extremely efficient and rapid depolarization owing to long-range electron-hole exchange interactions, which effectively couple excitons of opposite valley index \cite{glazov2014exciton,maialle1993exciton}.

\begin{figure}[h!]
	\center
	\includegraphics{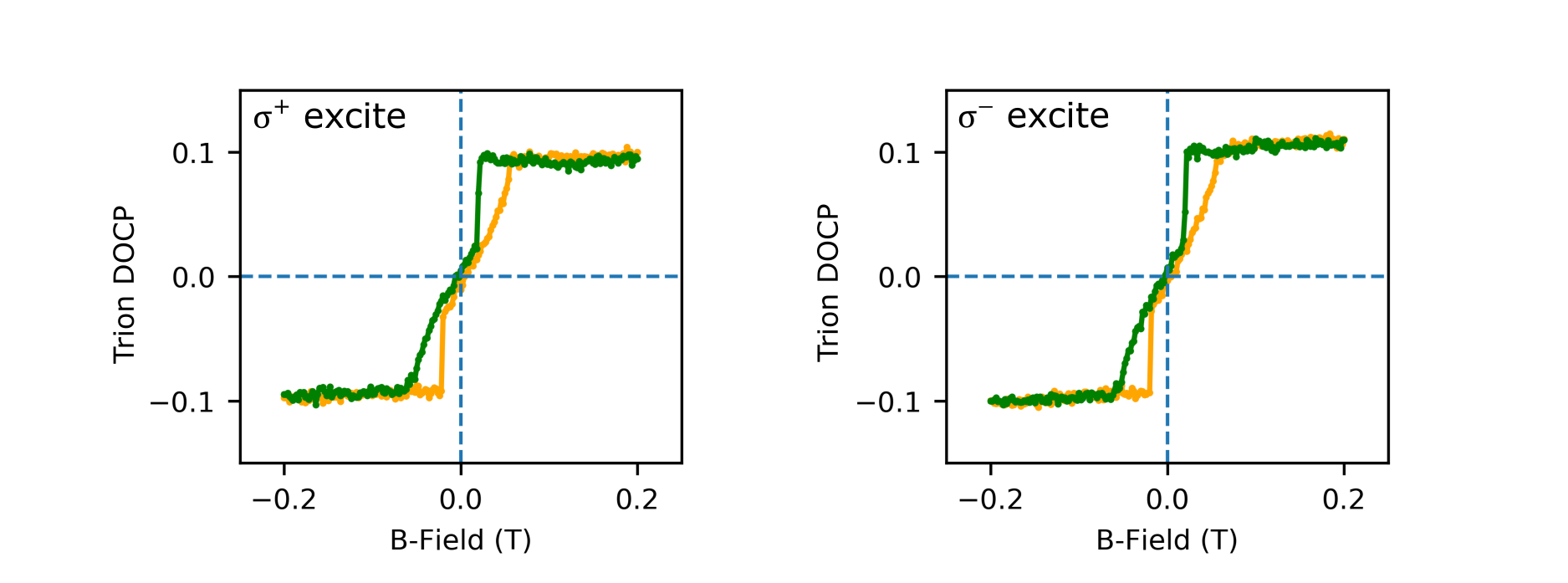}
	\caption{Comparison of the degree of circular polarization (DOCP, defined in the Main Text) for forward and backward B-field sweep directions, while under either $\sigma^+$ (left panel) or $\sigma^-$ (right panel) excitation laser polarization. The left panel is the same data as shown in the Main Text Fig. 2e. The identical result displayed here confirms that the polarization degree dependence is due to interactions with the CrBr$_3$ and external B-field, rather than the laser polarization, which has no effect owing to extremely rapid valley depolarization. This ensures that all optically generated valley polarization is lost before luminescence, if the excitation is sufficiently non-resonant, as is the case here (laser energy is $\sim300$ meV above the exciton energy).}
	\label{fig:laser_pol}
\end{figure}

To confirm that the polarization state of the laser in our experiments has no effect on the results, we repeat our experiments with $\sigma^-$ laser polarization, and measure polarization resolved PL intensity, as presented in the Main Text. The result for both laser polarizations are shown in Suppl. Fig.~\ref{fig:laser_pol}, where it is clear that the PL response is essentially identical regardless of laser polarization, confirming that any non-zero polarization in emission is a result of interaction with the ferromagnet or external B-field, and not the laser itself.

\clearpage

\subsection{Supplementary Note 6: Peak fitting for extraction of linewidth and valley splitting}

In order to extract the linewidths of the exciton and trion states as a function of applied external B-field (as shown in Main Text Fig. 3c), the polarization resolved photoluminescence spectra at each B-field increment were fitted to two Gaussian peak functions, corresponding to X and T. From the fitting we also extract a small valley splitting in both states, attributed to conduction band splitting as discussed in Supplementary Note 1. While the DFT calculations predict a larger proximity induced CB valley splitting of $\sim 2$ meV, the smaller experimental value may be due to the fact that interlayer exchange couplings are notoriously difficult to compute accurately,  the fact that lattice alignment may be different from the one assumed in the calculation, and the fact that excitonic effects have not been included in the calculation.

\begin{figure}[h!]
	\center
	\includegraphics[width=0.8\textwidth]{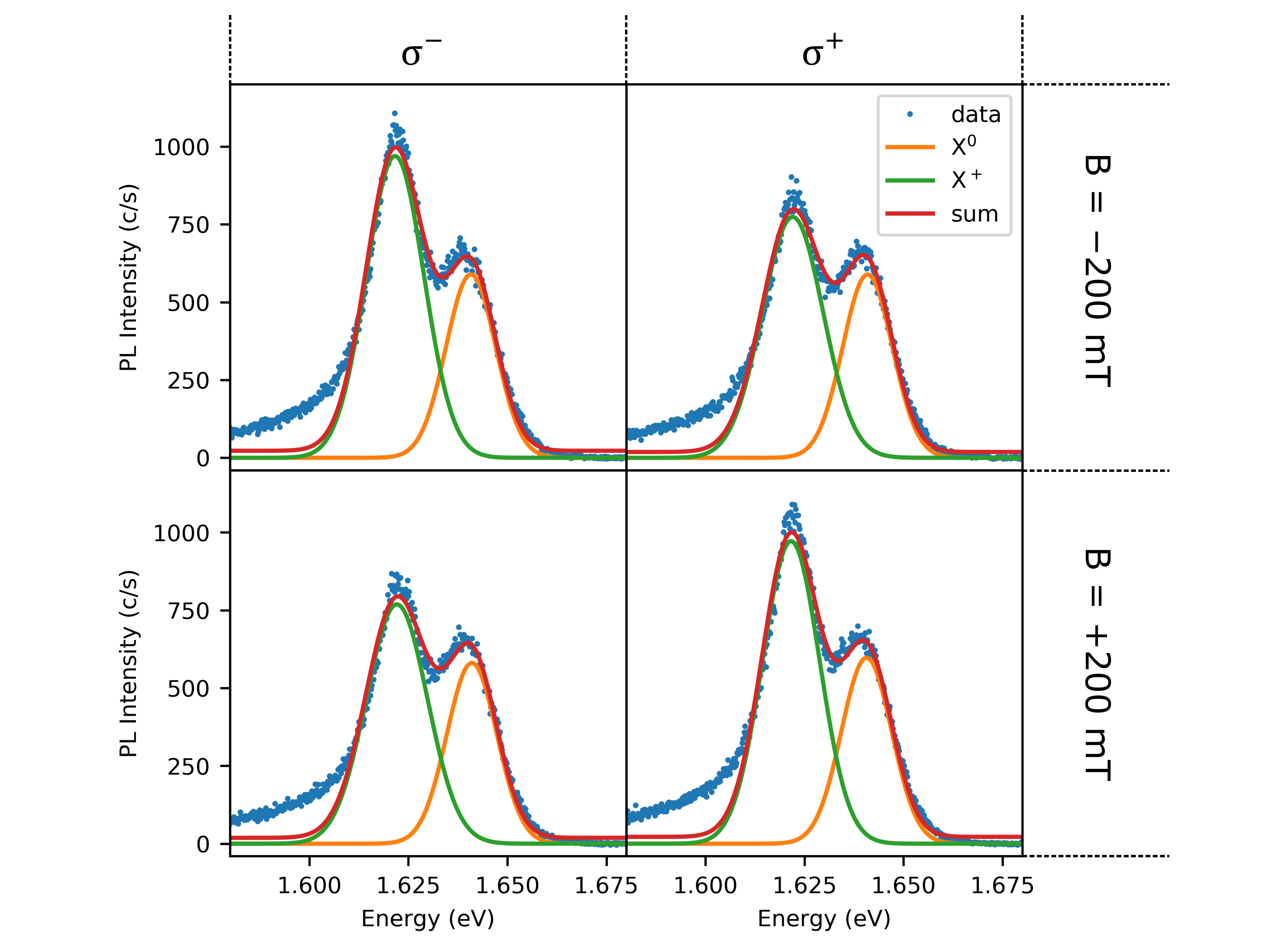}
	\caption{Examples of photoluminescence spectra in both $\sigma^+$ and $\sigma^-$ emission polarizations at external B-fields of $\pm200$ mT applied out of the sample plane. Each spectrum is fitted to two Gaussian peak functions.}
\end{figure}

\begin{figure}[h!]
	\center
	\includegraphics{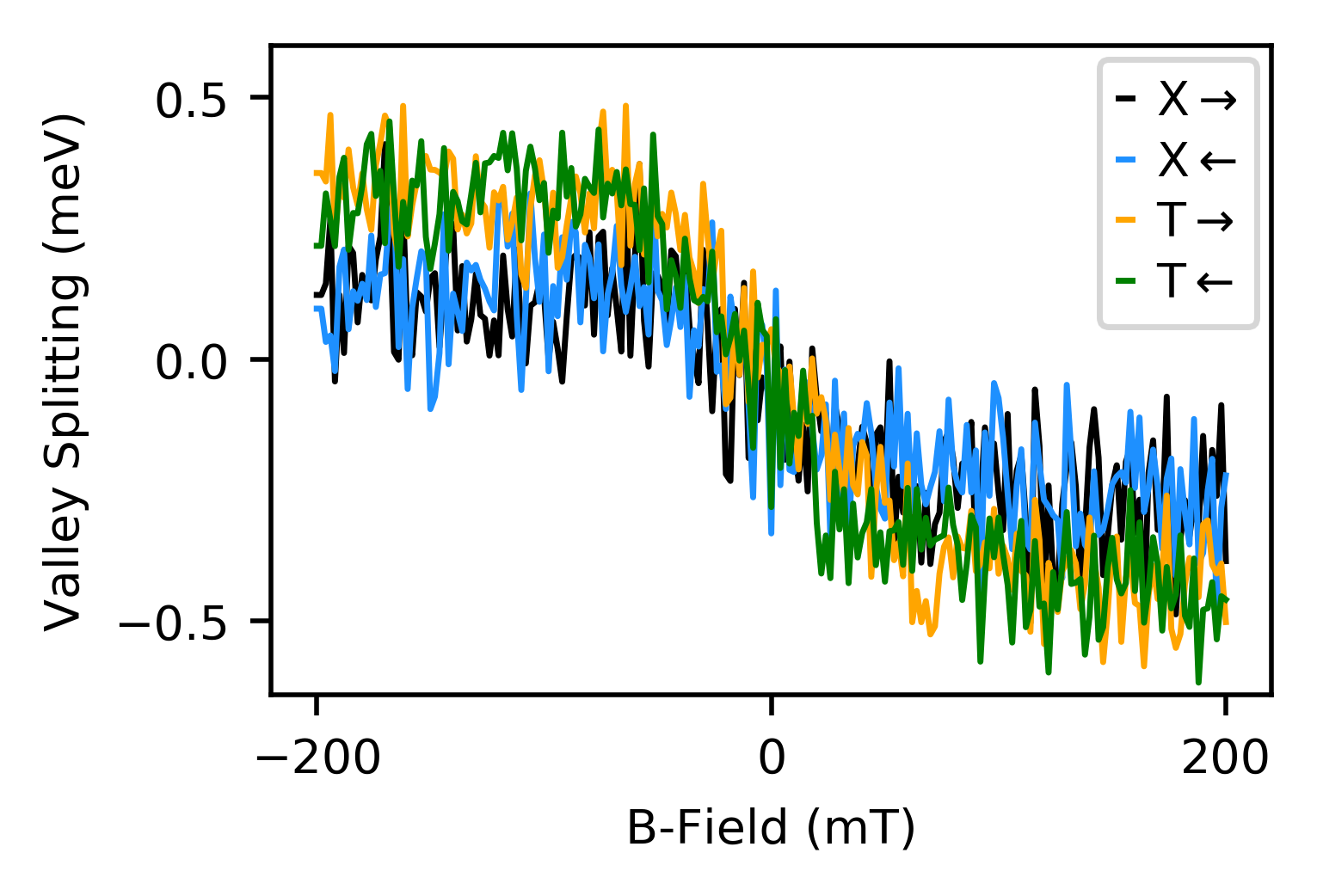}
	\caption{Valley splitting of exciton (X) and trion (T) PL peaks extracted from the Gaussian peak fitting (both sweep directions).}
\end{figure}

\clearpage

We also fit the same spectra with 2 peaks: a higher energy symmetric Gaussian function for the neutral exciton, and a lower energy peak for the trion, which is the convolution between a Gaussian and a low energy exponential function, in order to attempt to account for the trion tail. Examples of the alternative fitting are shown below, along with the extracted standard deviation (a direct measure of FWHM in the exponentially modified Gaussian is not a well defined parameter, and so we use standard deviation instead to display how the peak width varies with B) and valley splitting vs B-field. As can be seen, the results are qualitatively in agreement with the double Gaussian peak fitting method shown above. The trion width switches between two values while the exciton remains almost constant (in agreement with Main Text Fig. 3c), while both peaks show a small valley splitting (in agreement with the valley splitting shown on the previous page).

\begin{figure}[h!]
	\center
	\includegraphics[scale=0.9]{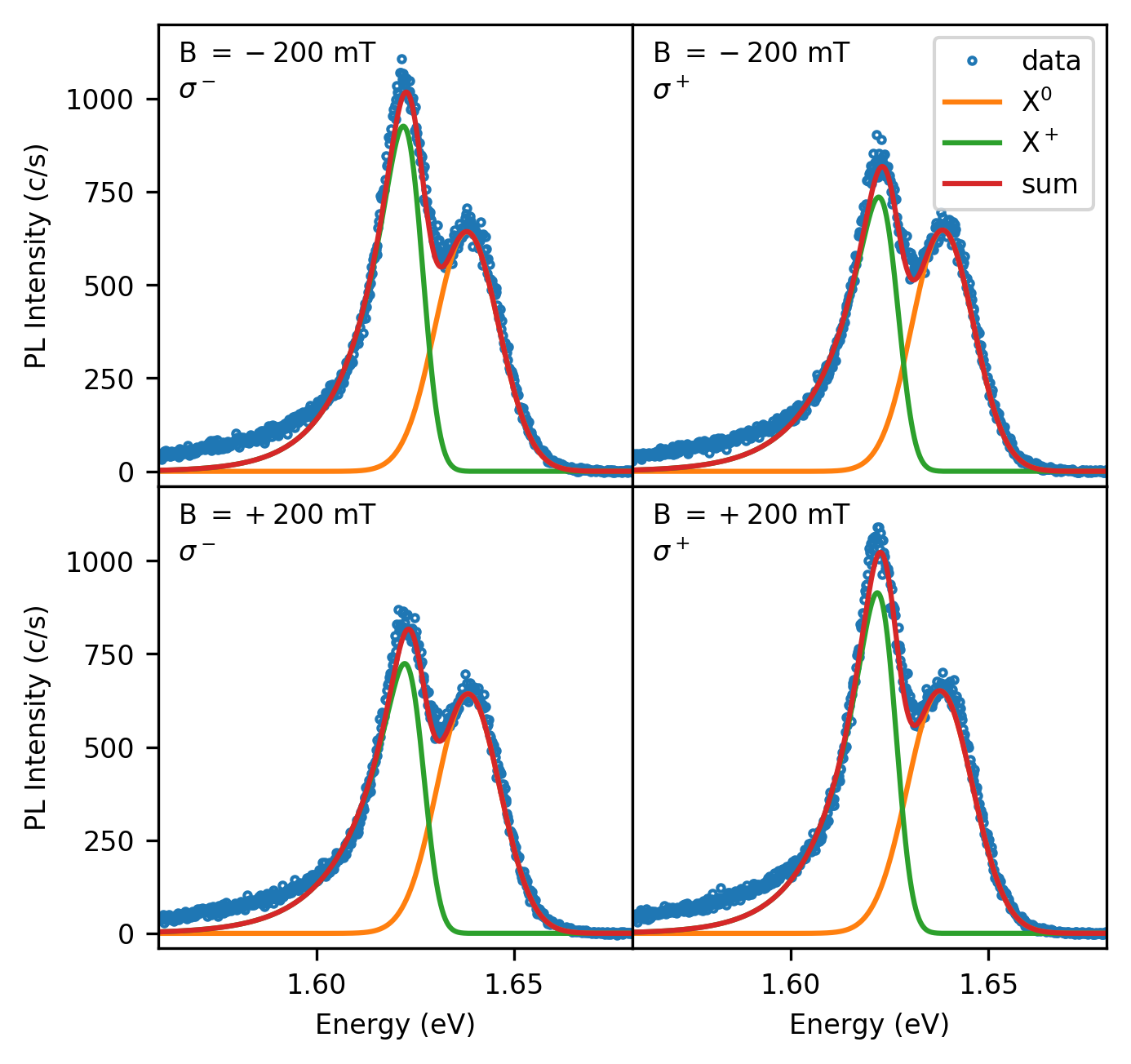}
	\caption{Examples of peak fitting the spectra using a Gaussian for the exciton and a Gaussian convolved with an exponential function for the trion.}
\end{figure}

\begin{figure}[h!]
	\center
	\includegraphics{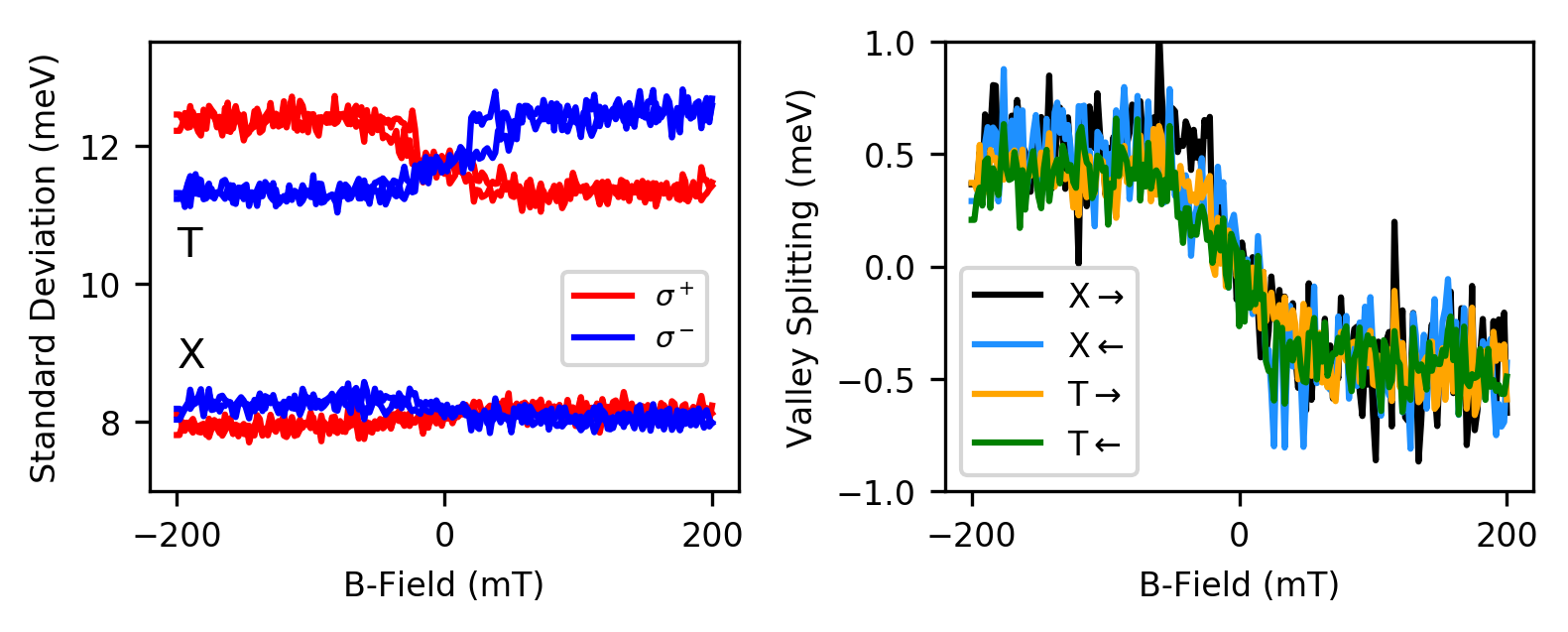}
	\caption{Standard deviation (while not exactly equal to the FWHM, this gives a measure of peak width) and valley splitting of exciton (X) and trion (T) PL peaks obtained using the alternative peak fitting method.}
\end{figure}

\clearpage

\subsection{Supplementary Note 7: The valley Zeeman effect}

Throughout the Main Text we do not exceed 200 mT applied external B-field strength. This is to ensure that the valley Zeeman effect in MoSe$_2$ can be neglected, and so any observations must result purely from a magnetic proximity effect. The weak B-field serves only to control the domain dynamics in CrBr$_3$, and is too weak to have any appreciable effect directly on band energies in MoSe$_2$.

However, in order to investigate whether the CrBr$_3$ substrate modifies the valley Zeeman response of MoSe$_2$, we measure the valley splitting from the sample in applied fields up to B = 8 T. The result is shown below, where we additionally perform a linear fitting for both exciton and trion lineshifts to extract an average rate of shift. We find the exciton to shift at $(-0.28 \pm 0.01)$ meV / T while the trion shifts at $(-0.11 \pm 0.01)$ meV / T. Variation in rates of shift in MoSe$_2$ have been reported to be a consequence of doping level \cite{li2014valley}. We also expect that CrBr$_3$, although saturated above B = 200 mT, may also modify the absolute rates of shift. In general, the exact nature of interplay between band shifts due to the valley Zeeman effect, and those due to interfacial exchange field, are not well understood.

\begin{figure}[h!]
	\center
	\includegraphics{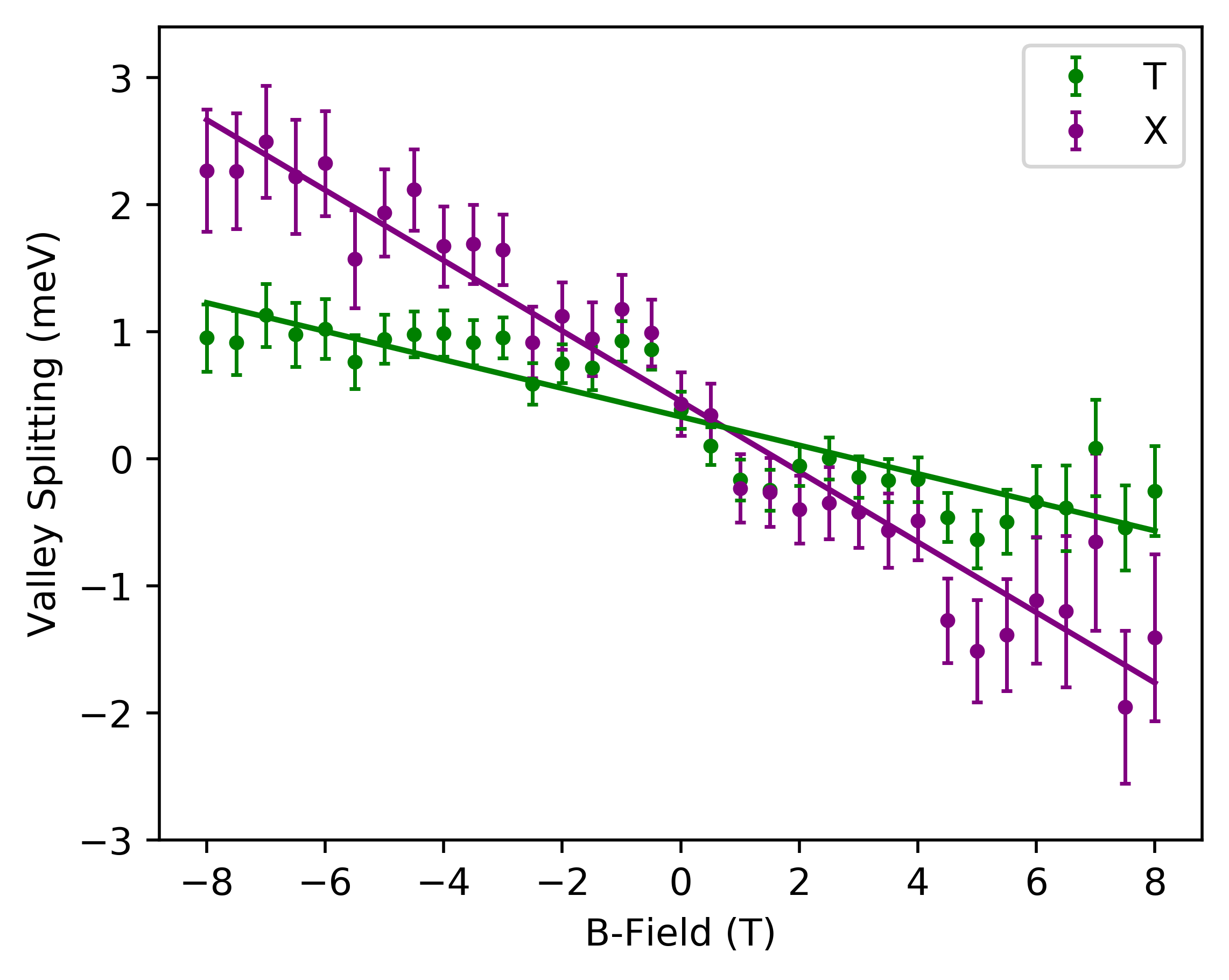}
	\caption{Valley splitting of exciton (X) and trion (T) PL peaks in external magnetic fields up to B = 8 T. Solid lines are linear fits to the data.}
\end{figure}

\clearpage

\subsection{Supplementary Note 8: Low temperature wide field Kerr microscopy}

In addition to the magneto-photoluminescence measurements presented in the Main Text, the sample was studied using a low temperature wide-field Kerr microscope. Details of the wide field Kerr microscopy are given in the methods section. Despite storage under high vacuum and dark surroundings, significant degradation was observed to have occurred in the CrBr$_3$ flake in the period between completing the PL measurements and commencing the Kerr measurements. This most likely happened during sample transit in low vacuum, and during periods of sample mounting and unmounting from cryostats in air. Figure 13a shows the sample before the PL measurements, with the CrBr$_3$ flake intact. Figures 13b and c show the sample before Kerr measurements, after degradation.

Figures 13d, e, and f show images of the sample from the Kerr microscope at out of plane magnetic field strengths of $B = -80, 0, +80$ mT, and a sample temperature of 11 K. A background image, taken at zero applied field, is subtracted from the live CMOS feed, such that any magnetization manifests as brighter or darker regions of the image relative to the non-magnetic surroundings. This occurs as light reflected from magnetized material displays a small rotation of the linear polarization plane, and so is transmitted through the analyzer by a greater or lesser extent. In this way, the spatially resolved brightness of the image denotes the polar Kerr signal.

The brightness change of the remaining CrBr$_3$ is very clear, becoming darker than the surroundings at negative field and brighter at positive field. The change in brightness of the surrounding substrate is due to Faraday rotation in the microscope objective and cryostat window. Crucially, no domain structures could be resolved in the Kerr microscope at any field strength, rather, the entire flake appears to brighten and darken at the same rate. This indicates that the domains are smaller than the optical resolution of the microscope, $\sim 300$ nm. If they were larger, a mixed pattern of bright and dark regions would be visible in Figure 9e at zero field, but no such pattern is observed.

\begin{figure}[h!]
	\center
	\includegraphics{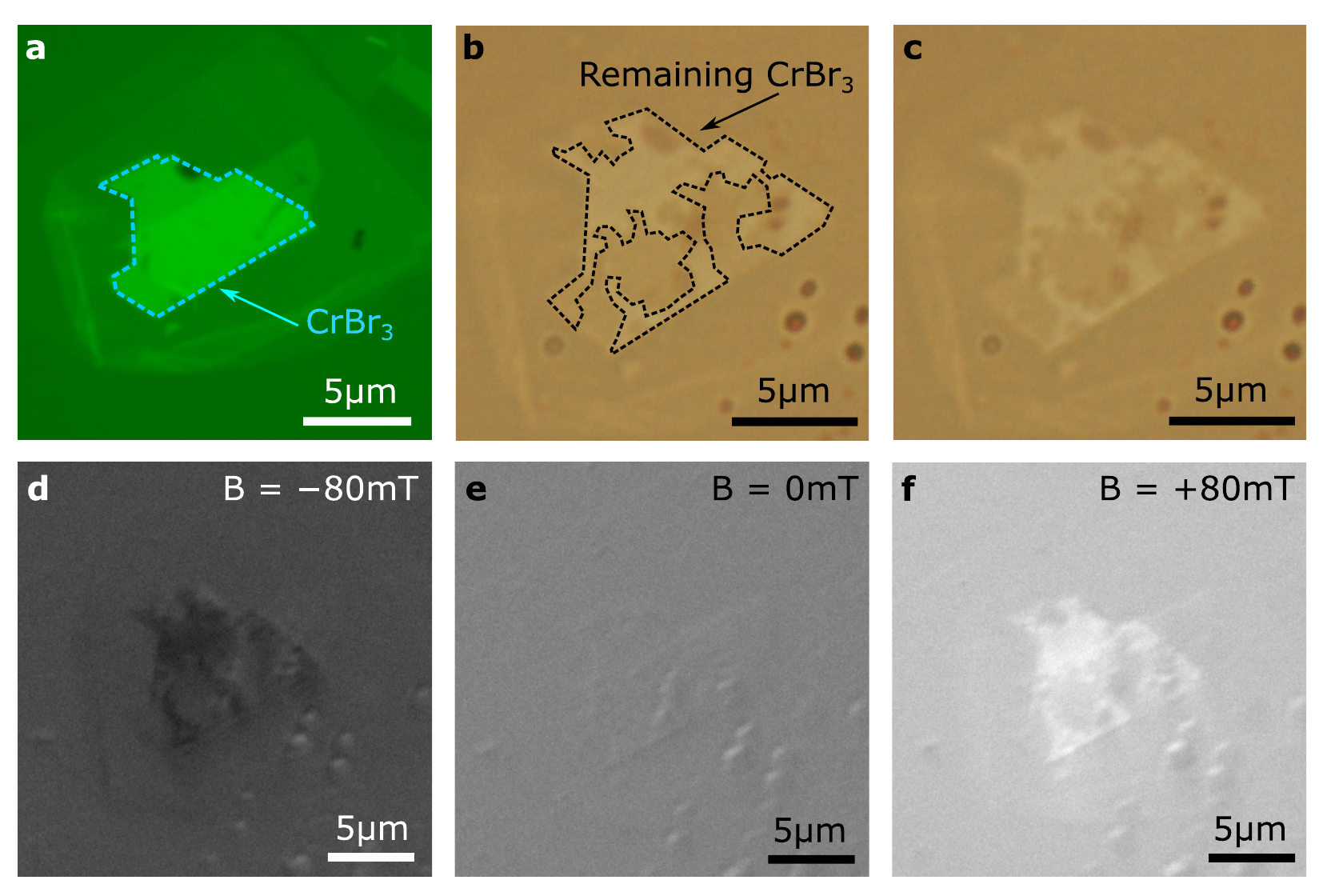}
	\caption{Sample degradation and wide field Kerr microscopy. (a-c) Bright field microscope images of the sample (a) before the PL measurements presented in the Main Text, (b) before Kerr measurements, showing degradation of the flake, and (c) as (b) but without annotations for clarity. (d) Kerr microscope image of the sample under a magnetic field of B $= -80$ mT out of the plane. The dark CrBr$_3$ coresponds to a magnetization aligned to B. (e) Kerr microscope image at zero field. No domain structures can be resolved. (f) Kerr microscope image at B $=+80$ mT. The brightness of the flake indicates magnetization aligned to B. All Kerr microscopy is performed at a sample temperature of 11 K.}
\end{figure}

\clearpage

\bibliographystyle{apsrev4-1_custom}
\bibliography{biblio_supp}